# Highly luminescent silver-based MOFs: Scalable eco-friendly synthesis paving the way for photonics sensors and electroluminescent devices

*Mario Gutiérrez, Cristina Martín, Barbara E. Souza, Mark Van der Auweraer, Johan Hofkens, and Jin-Chong Tan\**

Dr. M. Gutiérrez, Barbara E. Souza, Prof. J.C. Tan
[1]Multifunctional Materials & Composites (MMC) Laboratory, Department of Engineering Science, University of Oxford, Parks Road, Oxford OX1 3PJ, United Kingdom.

*Corresponding author
E-mail: jin-chong.tan@eng.ox.ac.uk

Dr. C. Martín
[2]Unidad de Medicina Molecular, Centro Regional de Investigaciones Biomédicas, Albacete, Spain.
[3]Molecular Imaging and Photonics, Department of Chemistry, Katholieke Universiteit Leuven, Celestijnenlaan 200F, 3001 Leuven, Belgium.

Prof. M. Van der Auweraer, Prof. J. Hofkens
[3]Molecular Imaging and Photonics, Department of Chemistry, Katholieke Universiteit Leuven, Celestijnenlaan 200F, 3001 Leuven, Belgium.

## Abstract

Luminescent multifunctional nanomaterials are important because of their potential impact on the development of key technologies such as smart luminescent sensors and solid-state lightings. To be technologically viable, the luminescent material needs to fulfil a number of requirements such as facile and cost-effective fabrication, a high quantum yield, structural robustness, and long-term material stability. To achieve these requirements, an eco-friendly and scalable synthesis of a highly photoluminescent, multistimuli responsive and electroluminescent silver-based metal-organic framework (Ag-MOF), termed "OX-2" was developed. Its exceptional photophysical and mechanically resilient properties that can be reversibly switched by temperature and pressure make this material stood out over other competing luminescent materials. The potential use of OX-2 MOF as a good electroluminescent material was tested by constructing a proof-of-concept MOF-LED (light emitting diode) device, further contributing to the rare examples of electroluminescent MOFs. The results reveal the huge potential for exploiting the Ag-MOF as a multitasking platform to engineer innovative photonic technologies.

## 1. Introduction

The development of luminescent smart nanomaterials is at the forefront of technological innovations given their many potential applications, including photonics sensors, visible light communications, anti-counterfeiting and bioimaging technologies [1-4]. Amongst those applications, luminescent nanomaterials that are highly responsive to external stimuli have gained much interests because they can function as non-invasive sensing devices (e.g. highly sensitive sensors of pollutants, volatile organic compounds (VOCs), pH, pressure or temperature) [5-8], while electroluminescent nanomaterials can function as the emissive layers in energy-efficient light-emitting diodes (LEDs) [9,10].

Metal-organic frameworks (MOFs), a class of crystalline hybrid compounds with versatile physicochemical properties, have emerged as one of the most prominent materials over the last two decades for the aforementioned applications [11,12]. Particularly, the field of luminescent MOFs (known as LMOFs) has been boosted over the last years through the development of numerous new examples of LMOF materials, where their chemical structure and physical properties can be adapted to afford different applications such as chemical sensors (i.e. VOCs, nitroaromatics, or antibiotic sensing), lighting, luminescent thermometry, and bioimaging [13-15]. For example, a luminescent two-dimensional MOF nanosheets material (NTU-9-NS) has been shown to be an excellent candidate for detecting small quantities (few ppm) of $Fe^{3+}$ ions, pertinent to environmental sustainability and biological applications [16]. Other examples concerned dye@MOF composites whose luminescent sensing response can be harnessed for the discrimination of a variety of VOCs [17,18]. Even though all those fields above involved some examples of certain LMOF materials, the enormous potential of tuneable LMOFs with multitasking sensing capabilities [19] is not yet well explored.

There exists an increasing demand on the development of a new generation of luminescent thermometers by using LMOFs, which can overcome the intrinsic limitations of conventional systems [13,20,21]. For example, disadvantages of the commercially available thermometers like fragility, slow response since they operate in contact mode, lacking high accuracy and high sensitivity subject to electric and magnetic fields [22,23], can be addressed by employing tuneable LMOFs materials. Besides these drawbacks, the development of nano-sized LMOFs luminescent thermometers does not only reduce the existing limitations, but they are opening the door to a whole range of possibilities, such as for understanding heat transfer mechanisms occurring at living cells or in integrated electronic circuits as they offer a fast response time, non-invasive operation (no surface contact), and they are not affected by electric and magnetic fields [24-26]. While recognizing that thermochromic LMOFs are just beginning to be explored, most examples are relying on expensive lanthanides and non-environmentally friendly rare earths, and/or requiring complicated synthetic methodologies thus hindering real world applications [13,20,21]. For instance, some representative examples include luminescent mixed Eu-Tb MOFs, which exhibit a ratiometric response to changes in temperature by leveraging the energy transfer mechanism from $Tb^{3+}$ to $Eu^{3+}$ [27,28].

Another big challenge that scientists is starting to face with LMOFs is the fabrication of new energy-efficient LEDs that are free of rare earths. Hitherto, although many LMOFs have been reported, there are few examples of LMOFs applied to solid-state lightings [29], and the number of studies is even smaller when considering only the electroluminescent MOFs functioning as the emissive layers of new LEDs [30-32]. Hitherto, those examples are reduced to the NNU-27 ($[ZnNa_2(L)_2(DEF)_2]\cdot DEF$) MOF where the LED emits orange-red light [30], an Sr-based MOF ($Sr(ntca)(H_2O)_2\cdot H_2O$, where ntca = 1,4,5,8-naphthalenetetracarboxylic acid hydrate) and a Zr-NDC (NDC = 2,6-naphthalenedicarboxylic acid) MOF where the LEDs produce white light [31,32]. Finally, a very recent alternative approach comprising the encapsulation of an electroluminescent guest into the ZIF-8 MOF has been developed, giving rise to the fabrication of an orange/yellowish light-emitting LED [33].

To address the foregoing challenges in the field of LMOFs, and with the aim of addressing several photonic technologies at once, herein we report for the first time a rapid eco-friendly synthesis of a highly photoluminescent, multistimuli-responsive and electroluminescent silver-based MOF: Ag-BDC (BDC = 1,4-benzene dicarboxylate), thereafter designated as "OX-2" (i.e. Oxford University-2 material). The selection of Ag as the metal cluster of the framework is motivated by the following reasons: i) its biocompatibility and low toxicity [34], ii) its affordability compared with rare-earth elements, and iii) its outstanding luminescence and semiconducting properties. Together, the foregoing factors make Ag-based luminescent materials a promising candidate for fabricating sustainable and high performance LEDs [35,36]. Based on that, we elucidate four synthetic approaches employing different processing parameters, where we scale up the reaction under ambient conditions to achieve large amounts of highly crystalline OX-2 using water as the solvent. While the crystalline structure of OX-2 is unaffected by the synthetic methodology, we show the luminescent properties and especially the quantum yield (ranging from 1% to 60%) is strongly affected by the coordinated solvent and the formation of new substructures in one of the synthetic procedures. A similar Ag-MOF structure to that of OX-2 was previously reported [37], however the more intricate and slower (one week) synthetic procedure as well as the lack of information about its photoluminescent or electroluminescent properties, prevent the direct comparison of the performance between both materials. It is worth noting that besides the promising optoelectronic properties of OX-2, which can be used for LED lighting applications, the remarkable water and air stability of OX-2 (for at least 21 and 70 days, respectively), along with its linear, repeatable and reproducible thermochromic response bode well for application as a luminescent thermometer.

## 2. Materials and Methods

### 2.1 Materials

Terephthalic acid (BDC, 99%+), $AgNO_3$, triethylamine ($NEt_3$, 99%), *N, N*-dimethylformamide (DMF, 99%) and methanol were purchased from Fisher Scientific and used without further purification.

### 2.2 Synthesis of OX-2 MOFs

#### *2.2.1 Synthesis 1: methanol as solvent (OX-2$_m$)*

6.0 mmol of BDC were deprotonated in a solution of 12.0 mmol of $NEt_3$ in 20 mL of methanol. Another solution was prepared by sonicating 3.0 mmol of $AgNO_3$ in 20 mL of methanol. The latter solution was added to the former one and a white suspension was promptly formed. The sample was sonicated for 5 minutes and then washed two times with methanol. The white solid sample was collected by centrifugation (8000 rpm) and dried at 80 °C for 2 hours. This procedure gave 430 mg of white OX-2 powder.

#### *2.2.2 Synthesis 2: Different reactants proportion in methanol (OX-2$_{m:1/2}$)*

Similar procedure to synthesis 1 was followed, but in this case, the amount (in mmol) of BDC was reduced to a half compared to $AgNO_3$. Briefly, 1.5 mmol of BDC and 3 mmol of $NEt_3$ were dissolved in 20 mL of methanol and this solution was added to a 3.0 mmol $AgNO_3$ solution in 20 mL of methanol. The mixture was sonicated for 5 minutes, washed with methanol and dried at 80 °C for 2 hours. 480 mg of a brownish powder of OX-2 was obtained.

#### *2.2.3 Synthesis 3: N, N-dimethylformamide as solvent (OX-2$_{DMF}$)*

Similar methodology to synthesis 1 was adapted using *N, N*-dimethylformamide (DMF) instead of methanol as solvent. Typically, 6.0 mmol of BDC and 12.0 mmol of $NEt_3$ were dissolved in 20 mL of DMF and then added to another solution prepared by sonicating 3.0 mmol of AgNO$_3$ in 20 mL of DMF. The mixture was sonicated for 5 minutes and then washed with DMF and methanol, and dried at 80 °C for 2 hours. 420 mg of a white OX-2 powder were obtained.

### *2.2.4 Synthesis 4: water as solvent (OX-2$_w$)*

Similarly, 6.0 mmol of BDC and 12.0 mmol of NEt$_3$ were dissolved in 20 mL of deionized water and this solution was added to another prepared by dissolving 3.0 mmol of AgNO$_3$ in 20 mL of deionized water. The instantaneously formed white suspension of OX-2 was sonicated for 5 minutes, washed with deionized water, collected by centrifugation and dried at 100 °C for 3 hours. 450 mg of white OX-2 powder were obtained.

### *2.2.5 Scalable synthesis of OX-2 in water*

The above described procedure was adapted to scale-up the synthesis of OX-2. Briefly, 120 mmol of BDC and 240 mmol of NEt$_3$ were dissolved in 300 mL of deionized water. Another solution was prepared by dissolving 60 mmol of AgNO$_3$ in 300 mL of deionized water. After that, the latter solution was added to the former one and a white suspension of a highly luminescent OX-2 (green emission under 365 nm UV) was instantaneously formed (see video 1 in SI). The mixture was sonicated for 5 minutes, then washed with a copious amount of deionized water, collected by centrifugation and dried at 100 °C for 3 hours. Following this procedure, we could easily obtain 10 g of OX-2$_w$ MOF, proving the efficacy of the eco-friendly scaled-up synthesis.

### **2.3 Materials Characterization**

The crystalline structure, morphology and luminescent properties of OX-2 MOFs were characterized by a combination of X-ray, spectroscopy and microscopy techniques. Powder X-ray diffraction (PXRD) was carried out in a Rigaku Miniflex diffractometer with a Cu Kα source (1.541 Å). The diffraction data were collected using 0.01° step size, 1°/min and for 2$\theta$ angle ranging from 2° to 32°. Field emission scanning electron microscopy (FESEM) and energy-dispersive X-ray (EDX) images and spectra were obtained using the Carl Zeiss Merlin FESEM equipped with a high-resolution field emission gun. Micrographs were attained in secondary electron imaging mode under high vacuum with an accelerating voltage of 10 keV.

FTIR spectra were recorded on a Nicolet iS10 spectrometer. The FTIR spectrum of each sample was collected 3 times and then averaged. Steady-state fluorescence spectra, excitation-emission maps, luminescence quantum yields and time-resolved emission decays were recorded using the FS-5 spectrofluorometer (Edinburgh Instruments) equipped with different modules for each specific experiment (i.e. integrating sphere for quantum yield, heated sample module to measure the emission of powders at different temperatures and a standard solid holder for powder experiments). For time-resolved fluorescence lifetime measurements, the samples were pumped with a 365-nm centred pulsed diode laser. The instrumental response function (IRF, ~800 ps) was used to deconvolute the emission decays. The decays were fitted to a multiexponential function and the quality of the fit was estimated by the $\chi^2$, which was always below 1.2.

**2.4 Light-Emitting Diodes (LEDs) Fabrication**

For the MOF-LED fabrication an indium-tin-oxide (ITO) coated glass substrate with a sheet resistance of ≈$10^9$ Ω cm was used as a substrate. Prior to use, the substrates were cleaned following the procedure described elsewhere [32]. Then, a solution of PEDOT:PSS (Sigma-Aldrich, high conductivity), which acts as the hole injection layer, was spin coated at 3000 rpm for 90 s and subsequently annealed at 140 °C for 10 min. After that, a well-dispersed suspension of OX-2 in chlorobenzene (10 mg/mL) or a dispersion of 10 mg of OX-2 MOF in a solution of poly(9-vinylcarbazole) polymer (PVK: MW ≈ 25,000–50,000 g/mol) in chlorobenzene (Sigma-Aldrich, anhydrous), was spin-coated at 1000 rpm for 60 s and annealed for 10 min at 80 °C. Next, a 5 mg/mL solution of the electron injection layer, 2-(4-tert-butylphenyl)-5-(4-

biphenylyl)-1,3,4-oxadiazole (PBD, Sigma Aldrich) in cyclohexane, was spin coated at 6000 rpm for 90 s and the layer was annealed at 80 °C for 10 min. Finally, a 150-nm thick layer of the aluminium (Al) electrode was vapour deposited on the top of the PBD layer.

## 3. Results and Discussion

### 3.1 Materials characterization

In this study, we have developed four different approaches for obtaining a luminescent Ag-based MOF material, termed OX-2. The synthetic conditions mainly differ in the solvent type used for the reaction and the ratio of reactants. To unravel how the different parameters affect the physicochemical characteristics of the resultant materials, their crystalline structure, morphology, and spectroscopic properties were determined by means of PXRD, FESEM-EDX, FTIR, and steady-state and time-resolved fluorescence spectroscopy.

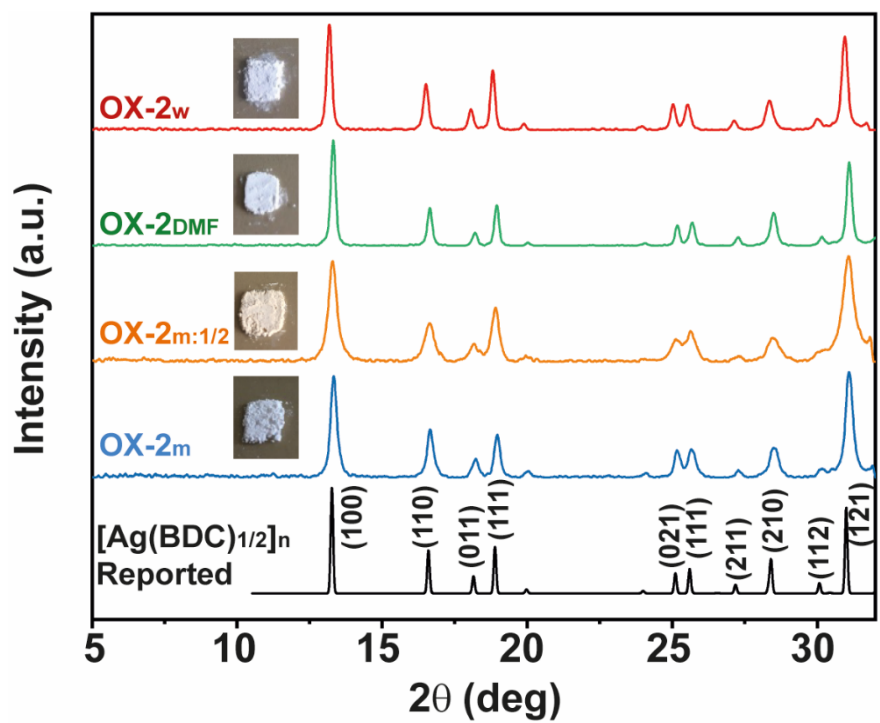

**Figure 1.** PXRD patterns of the $[Ag(BDC)_{1/2}]_n$ reported (CCDC 198096) and the different OX-2 MOFs synthesized in methanol (OX-2$_m$), in methanol but using a 1:2 ratio of BDC organic linker to AgNO$_3$ metal salt (OX-2$_{m:1/2}$), in DMF (OX-2$_{DMF}$), and in water (OX-2$_w$). The insets are photos of each powder sample showing the white colour of all materials with the exception of the brownish OX-2$_{m:1/2}$ sample.

We begin by examining the crystalline structures of the four OX-2 MOFs by comparing the PXRD patterns shown in **Figure 1**. It can be seen that the diffraction patterns of OX-2$_m$, OX-2$_{m:1/2}$, OX-2$_{DMF}$, and OX-2$_w$ are very similar. Specifically, the Bragg peaks revealed no significant changes with regards to the peak intensity and position. We established that the XRD patterns are resembling a previously reported Ag-carboxylate coordination polymer [37]. The molecular formula of this Ag network corresponds to [Ag(BDC)$_{1/2}$]$_n$ and the crystalline structure reveals a dense 3D framework with a short Ag—Ag bond distance of 2.901 Å. Basically, the binuclear Ag$_2$(BDC)$_2$ may be considered as the basic building units, where the Ag ions are linked through two bridging carboxylate groups of the BDC linkers (Scheme S1A). These building units are connected head-to-tail through the BDC linkers giving rise to 1D chains, which are connected to each other by secondary BDC linkers through weak Ag-O bonds, generating 2D wave-like layers. Finally, those 2D layers are interwoven to form a dense 3D framework (Scheme S1B) [37].

The FTIR spectra of the different OX-2 MOFs depicted in **Figure S1** show no remarkable differences apart from a broad band at ~1700 cm$^{-1}$ for OX-2$_{DMF}$, which can be attributed to the $\upsilon$(C=O) stretching mode of residual DMF molecules in OX-2 [38]. The FTIR spectra of all OX-2 samples exhibit vibrational bands at ~1520, 1360, 1300, 1150, 1090, 1010, 890, 820 and 740 cm$^{-1}$. Following previous assignments, the bands at 1520 and 1360 cm$^{-1}$ can be ascribed to asymmetric and symmetric stretching modes of the carboxylic groups of the BDC linker coordinated to the Ag metal centre [39]. Moreover, the lack of bands in the region between 1750-1680 cm$^{-1}$ confirmed a complete deprotonation of terephthalic acid. Finally, the bands observed in the region between 1150-740 cm$^{-1}$ are attributable to the $\upsilon$(C=C) stretching mode, $\beta$(CCH) and $\gamma$(CCC) bending modes of the benzene rings [40].

Although no significant structural variations and FTIR response have been detected for the different OX-2 materials, simple visual inspection reveals notable changes both under daylight and UV exposures. OX-2$_m$, OX-2$_{DMF}$ and OX-2$_w$ MOFs are purely white powders (see photo insets in **Figure 1**), however, under UV irradiation there are significant differences in the emission of these powders, whose intensity is declining in the order of OX-2 synthesized in water > methanol > DMF. In contrast, OX-2$_{m:1/2}$ is brownish in colour under daylight (see photo in **Figure 1**). Different explanations can justify this effect, for instance, as the aggregation of the undissolved silver to form nanoparticles will give a brownish colour [41]. However, it is also possible that the different ratio of silver nitrate and BDC linker used in synthesis could yield dissimilar MOF crystal morphology in OX-2. Subsequently, we performed FESEM characterization to determine the most plausible factor underpinning this effect.

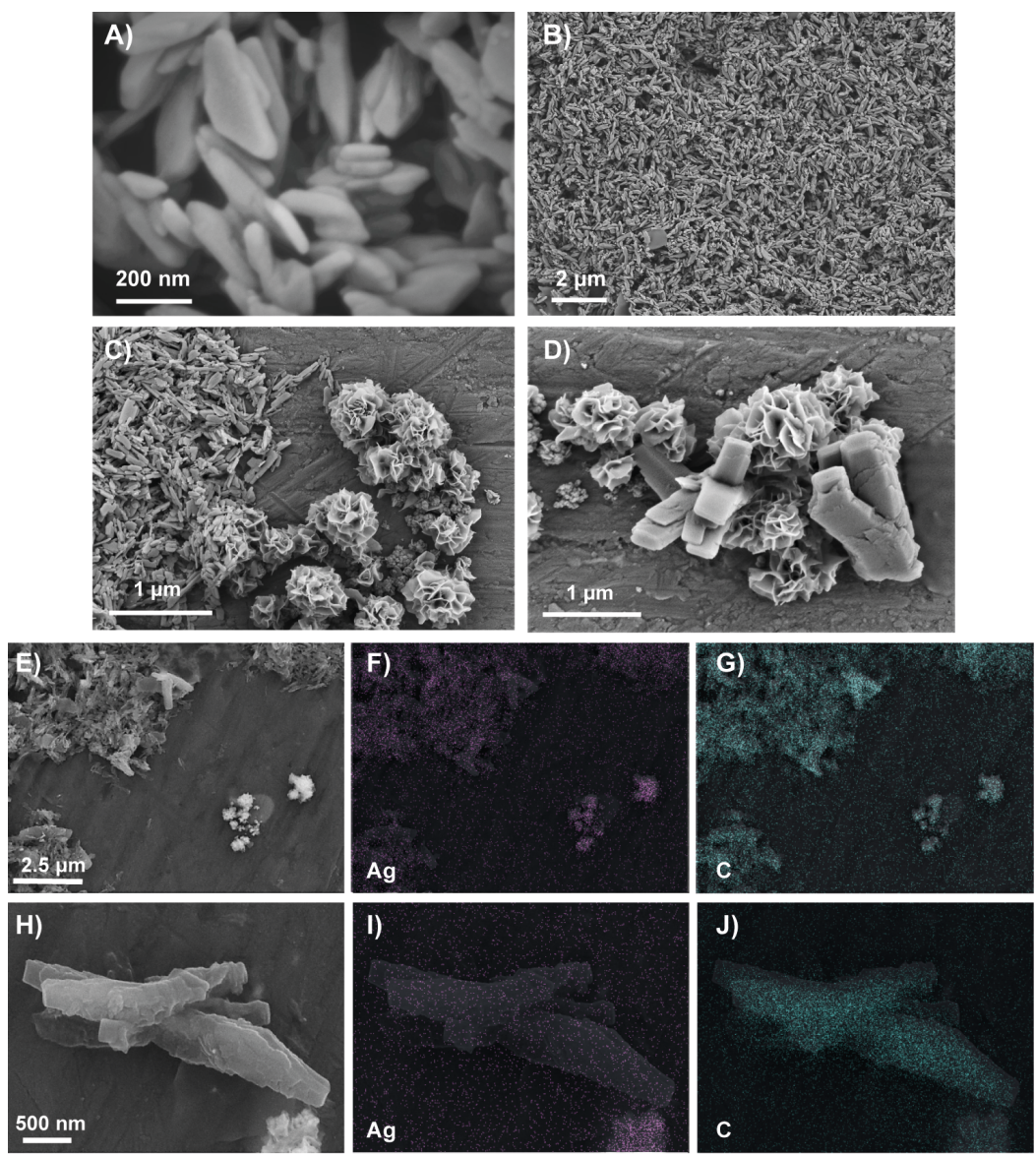

**Figure 2.** FESEM micrographs of **(A-B)** OX-2$_w$ and **(C-D)** OX-2$_{m:1/2}$. **(E-J)** FESEM-EDX analyses of the different crystal morphologies found in OX-2$_{m:1/2}$ MOF, showing the distribution of the elements of Ag (pink dots) and C (green dots) identified in the representative crystals.

**Figures 2A-D** and **Figure S2** display the micrographs of OX-2$_w$, OX-2$_{m:1/2}$, OX2$_m$ and OX-2$_{DMF}$, where appreciable morphological changes were observed. While the FESEM images of OX-2$_w$ (**Figures 2A-B**), OX2$_m$ and OX-2$_{DMF}$ (**Figure S2**) show a homogenous distribution of elongated nanoplates (height 60-100 nm, **Figure S3**), OX-2$_{m:1/2}$ presents three different morphologies: (i) elongated nanoplates, (ii) aggregation of a second type of nanoplates with "flower"-shaped crystals, and (iii) micron-sized "columns" (**Figures 2C-D**). To get more insights into the chemical composition of these three different morphologies, FESEM-EDX mapping of the OX-2$_{m:1/2}$ and OX-2$_w$ MOFs was performed (**Figures 2E-G**, **Figures S4A-C** and **Figures S4D-F**, respectively) as they exhibit the most dissimilar characteristics (white and highly luminescent versus brownish and almost non-emissive powders, respectively). Interestingly, whereas the elemental composition of the elongated nanoplates and flowers indicates the presence of uniformly distributed Ag, C and O atoms, the crystals with a columnar morphology are mainly comprising C and O atoms with no traces of Ag detected (apart from background signals). Moreover, the elongated nanoplates and flower crystals also present important differences in the Ag/C atomic ratio, being 1/16 and 1/30, respectively (**Table S1**). Therefore, the EDX results reveal the differences are attributed to the formation of additional substructures in OX-2 MOF, instead of silver nanoparticle formation. This is further confirmed by the emission spectrum of our material (**Figure 3**) which completely differs from that of the silver nanoparticles reported [41]. Additionally, it is anticipated that the presence of the secondary substructures will have a negative impact on the photoluminescence properties (*vide infra*).

**3.2 Photoluminescent properties of OX-2 MOFs**

As shown in the insets of **Figure 3**, when the OX-2 MOFs were exposed to UV irradiation ($\lambda = 365$ nm) they displayed a strong green-yellow emission, with the exception of OX-2$_{m:1/2}$ which was barely emissive when contrasted against its counterparts. To better understand their luminescent properties, we have carried out steady-state excitation-emission mapping, fluorescence quantum yield, and time-resolved emission experiments. The excitation-emission maps of OX-2$_m$, OX-2$_{m:1/2}$, OX-2$_{DMF}$ and OX-2$_w$ reveal no significant differences, with all of them showing a vibrationally-resolved emission, having maxima located at 485, 520 and 560 nm that correspond to an excitation maximum of ~330 nm (**Figure 3**). Although the emission spectra of all OX-2 MOFs are similar, there are large variations detected in the value of photoluminescence quantum yield ($QY_{Exc(330nm)}$), which changes from 63% > 46% > 15% > 1% corresponding to the bulk powder samples of OX-2$_w$, OX-2$_m$, OX-2$_{DMF}$ and OX-2$_{m:1/2}$, respectively.

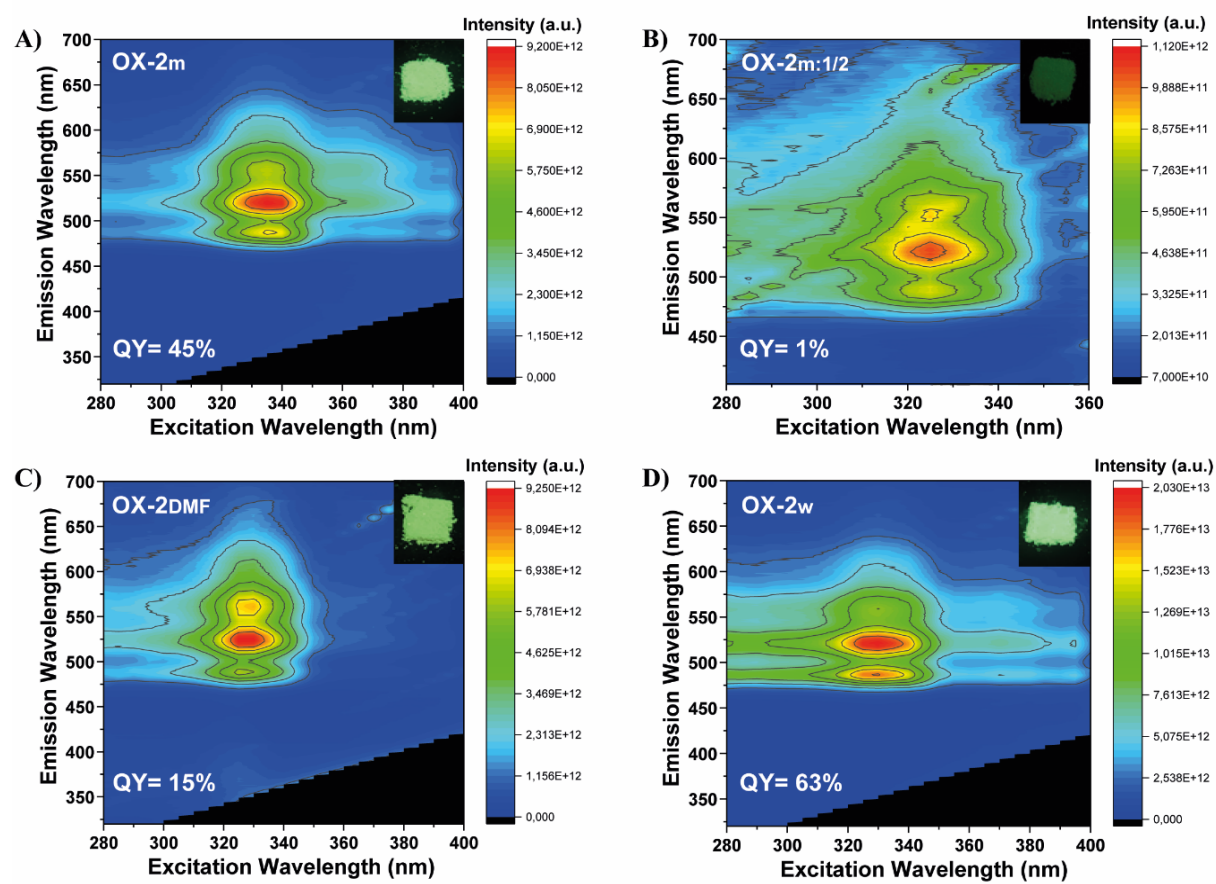

**Figure 3.** Excitation-emission maps of **A)** OX-2$_m$, **B)** OX-2$_{m:1/2}$, **C)** OX-2$_{DMF}$ and **D)** OX-2$_w$ recorded in solid-state. The quantum yield (QY) recorded using an excitation wavelength of 330 nm, along with the photo of each sample under UV irradiation (365 nm) are depicted as inset.

The large difference between the QY values of OX-2$_w$ and OX-2$_{m:1/2}$ is easily explained by the different crystalline morphologies found in the latter (*vide supra*). As described above, whereas OX-2$_w$ bulk powder consists of uniformly dispersed elongated nanoplates (**Figure 2B**), OX-2$_{m:1/2}$ is composed of a mixture of the flower- and columnar-shaped crystals in addition to the elongated nanoplates (**Figure 2C**). The elongated nanoplates of OX-2$_{m:1/2}$ are similar to those found in OX-2$_w$ and we can assume that those are the emissive microstructures. However, the flower-like morphology presents a different proportion of Ag/C while the columnar-shaped crystals are composed only by C and O atoms. Therefore, those species must be causing the decrease in the emission QY value of OX-2$_{m:1/2}$ mainly from two mechanisms: through the absorption of part of the photons from the irradiation source (brownish colour) and then deactivating by means of non-radiative transitions, and by absorbing part of the light emitted by the elongated nanoplates. For a more precise comparison, we have also synthesized the OX-2$_{w:1/2}$ in a similar way to OX-2$_{m:1/2}$, where the methanol was replaced by water. The obtained OX-2$_{w:1/2}$ is a brownish powder (under daylight) exhibiting a weak green emission under UV excitation, similarly to OX-2$_{m:1/2}$. The emission quantum yield of OX-2$_{w:1/2}$ is dramatically decreased to 8% when compared with the ~60% observed for OX-2$_w$. Therefore, it can be concluded that the ratio of reactants has a major impact on the emission properties, where the BDC:Ag ratio of 2:1 was found to be optimal.

More intriguing are the differences observed in the QY of OX-2$_w$, OX-2$_m$ and OX-2$_{DMF}$, as all of them are white solid powders with no morphological or structural differences. Therefore, if the synthesis of those MOFs was analogous and in all the cases the same structure and morphology were obtained (without any evidence of silver nanoparticles), the only interchangeable parameter, the solvent, is key to the differential emission of OX-2 MOFs. Recently it has been reported that the luminescence properties of Ag nanoclusters confined into zeolites strongly depends on the coordinated water molecules [42]. In fact, their optical properties have originated from a confined two-electron superatom quantum system *via* hybridization of Ag and water O orbitals, leading to a promotion of one electron from the *s*-type highest occupied molecular orbital (HOMO) to the *p*-type lowest unoccupied molecular orbital (LUMO) upon excitation [42]. This could explain why the quantum yield of OX-2$_w$ is the highest (63%) as water molecules might be coordinated to Ag atoms. Similarly, OX-2$_m$ also exhibits a high QY (46%) as the O atoms of the methanol solvent may be hybridized with the Ag atoms of the framework.

The emission of all OX-2 MOFs resembles that previously reported for the [Ag(L$^1$)]$_2$(p-BDC)·8H$_2$O coordination polymer with maxima at ~485, ~520 and ~560 nm [43]. In that study, the band at ~520 nm was assigned to a metal-to-ligand charge-transfer transition (MLCT), whereas the band at ~485 nm was attributed to intra-ligand emission as it was reminiscent of a weak emission at 466 nm previously reported for terephthalic acid [43]. However, this assumption is not fully supported by the following facts: (i) the intra-ligand emission of BDC is 20-nm blue shifted (466 nm) compared to that observed in OX-2 (485 nm). (ii) The intra-ligand BDC emission band is much broader (FWHM ~3250 cm$^{-1}$) [44] than the band at 485-nm observed in OX-2 MOFs (FWHM ~1050cm$^{-1}$, deconvoluted from OX-2 emission), whose narrowness matches better with a vibrationally-resolved structure, and (iii) it is possible to anticipate that the three vibrationally-resolved bands response in a comparable way to the external stimuli. On this basis, it is possible to rule out the band at 485-nm as the intra-ligand BDC emission. Instead, herein we propose that the vibrationally-resolved emission is attributed to a transition from a unique excited state, which is generated by a ligand-to-metal charge transfer (LMCT) and/or a ligand-to-metal-metal charge transfer (LMMCT) transitions. It has been extensively demonstrated that LMCT is one of the most plausible mechanisms for the luminescence of $d^{10}$ metal complexes [45]. Additionally, short metal-to-metal (M-M) distances may induce the emission from LMMCT transitions [46,47]. For instance, the emission of Ag-carboxylate nanoclusters has previously been attributed to LMMCT from Ag(I)-carboxylate complexes to Ag atoms [46]. Therefore, given that the Ag—Ag distance in OX-2 MOFs is only 2.901 Å apart, the possibility of luminescent transitions originating from LMMCT and/or LMCT transitions can be further substantiated.

To gather further evidence about the proposed phenomena, we performed time-correlated single photon counting (TCSPC) experiments on the four different OX-2 samples and the BDC linker, whose results are presented in **Figures S5 A-E** and **Tables S2-3**. Several important findings can be established from systematic analysis of the emission decay data. Firstly, while the BDC signal recorded at 485, 520 and 560 nm decays multiexponentially with time constants of $\tau_1$ = 1.7 ns, $\tau_2$ = 4.2 ns and $\tau_3$ = 16.5 ns, those of all OX-2 exhibit a biexponential behaviour with time constants of $\tau_1$ = 350-540 ps and $\tau_2$ = 2.0-3.5 ns (**Table S3**). This observation unequivocally proves that the band at 485-nm cannot be attributable to a BDC intra-ligand emission, and it also supports the notion that all the vibrationally-resolved bands share a common origin. Secondly, the decrease in the lifetimes observed for OX-2$_{m:1/2}$ compared with OX-2$_w$ ($\tau_1$ = 350 ps and $\tau_2$ = 2.0 vs $\tau_1$ = 540 ps and $\tau_2$ = 3.5 ns, respectively) agrees with our hypothesis that the flower- and columnar-shaped crystals are quenching the emission of the elongated nanoplates. Thirdly and more remarkably, upon the excitation of BDC linkers with UV light, the green-yellowish emission of OX-2 MOFs decays similarly to a previously reported Ag nanocrystals ($\tau_1 \sim$ 550 ps and $\tau_2 \sim$ 3.3 ns) [48], thus supporting the idea that the origin of the OX-2 luminescence is the LMCT and/or LMMCT transitions.

**3.3 Scalable synthesis and robustness of OX-2**

Encouraged by the easy synthesis and excellent photophysical properties of OX-2 MOF and in order to yield large quantities of this material for potential future applications, we developed a scale-up of the synthetic methodology described in section 2. Herein, we show that greater than 10 g of highly luminescent OX-2 material can be achieved simply by increasing the amount of initial reactants combined with water as solvent (see experimental section for details). Scaling up the synthetic protocol for MOF fabrication usually resulted in the formation of unwanted amorphous materials and reduced phase purity of the products, whose physicochemical properties are completely different from the crystalline product obtained in a small-scale reaction. Herein, our scalable approach using water has successfully yielded the identical OX-2$_w$, matching the phase synthesized in small scale as evidenced from the PXRD patterns (**Figure S6**) and luminescence properties with the expected vibrationally-resolved bands at 485, 520 and 560 nm (**Figure S7**), while maintaining the high QY$_{Exc(330nm)}$ value of ~60±3% in the solid state.

The potential applicability of a new technical material has to satisfy certain prescribed conditions, hence, the chemical stability of OX-2 in water and its long-term stability at ambient conditions has been tested. OX-2$_w$ was soaked in water for 1, 4, 8 and 21 days and then dried at 80 °C under vacuum for 2 hours. The stability of OX-2 was then confirmed by means of PXRD (**Figure S8**), which exhibits no variations in the peak position nor in the peak intensity, indicating that its crystalline structure is unaltered. Moreover, another sample of OX-2$_w$ was exposed to ambient conditions in the lab for up to 70 days (exposed to day light and ~40% humidity) and its stability was tested by PXRD and photoluminescent measurements (**Figures S9 and S10,** respectively). The PXRD patterns of OX-2$_w$ MOF show no changes upon its long-term exposure to ambient conditions (**Figure S9**). Similarly, the photoluminescence of OX-2$_w$ was systematically monitored by measuring the quantum yield values, whose minimal decrease from an initial value of 60% to 57% in 70 days (i.e. corresponding to a reduction of 5% in the normalized signal) has verified the high stability and structural resilience of the OX-2 MOF under ambient conditions for an extended period of time (**Figure S10**). Additionally, we have performed leaching experiments and confirmed a negligible degradation of OX-2$_w$ in water (see SI for more information).

**3.4 Luminescent OX-2 for Multistimuli Detection**

In this section we explore the luminescent response of OX-2 when subject to different external stimuli such as temperature or mechanical compression to evaluate its possible application as non-invasive sensors.

### *3.4.1 Mechanochromic response of OX-2: Compressive stress tests*

To investigate the effect of mechanical stress on the properties of OX-2, four batches of samples of 200 mg each were compressed into pellets under a nominal stress of 0.074, 0.148, 0.222 and 0.297 GPa, respectively, and further characterized by PXRD and fluorescence spectroscopy. As shown in **Figure 4A**, upon mechanical compression, the pellets exhibit notable changes in the intensity of the Bragg peaks. Specifically, there is a decrease in intensity of the peak at $2\theta$ ~13° corresponding to the (100) plane. By making a correlation between the two most intense peaks located at ~13° and ~31° (corresponding to the (100) and (121) planes, respectively), there is a clear decrease in the ratio with increasing pelleting stress (values presented in **Figure 4A**). However, even at high pressures, OX-2 pellets could retain their original crystalline structure as no significant shift of the peak positions was detected. Mechanical compression also induces important changes in the emission intensity of OX-2 pellets, which we found to translate into a variation of quantum yield values (**Figure 4B**). At lower pelleting stress (0.074 GPa), the emission intensity of the pellet is comparable to the as-synthesized powder of OX-2, with QY = 59% in both cases. However, upon increasing the pressure, there is a continuous fall in the emission intensity, and thus, causing a declining QY values from 57% > 47% > 45% for pelleting pressures corresponding to 0.148, 0.222 and 0.297 GPa, respectively.

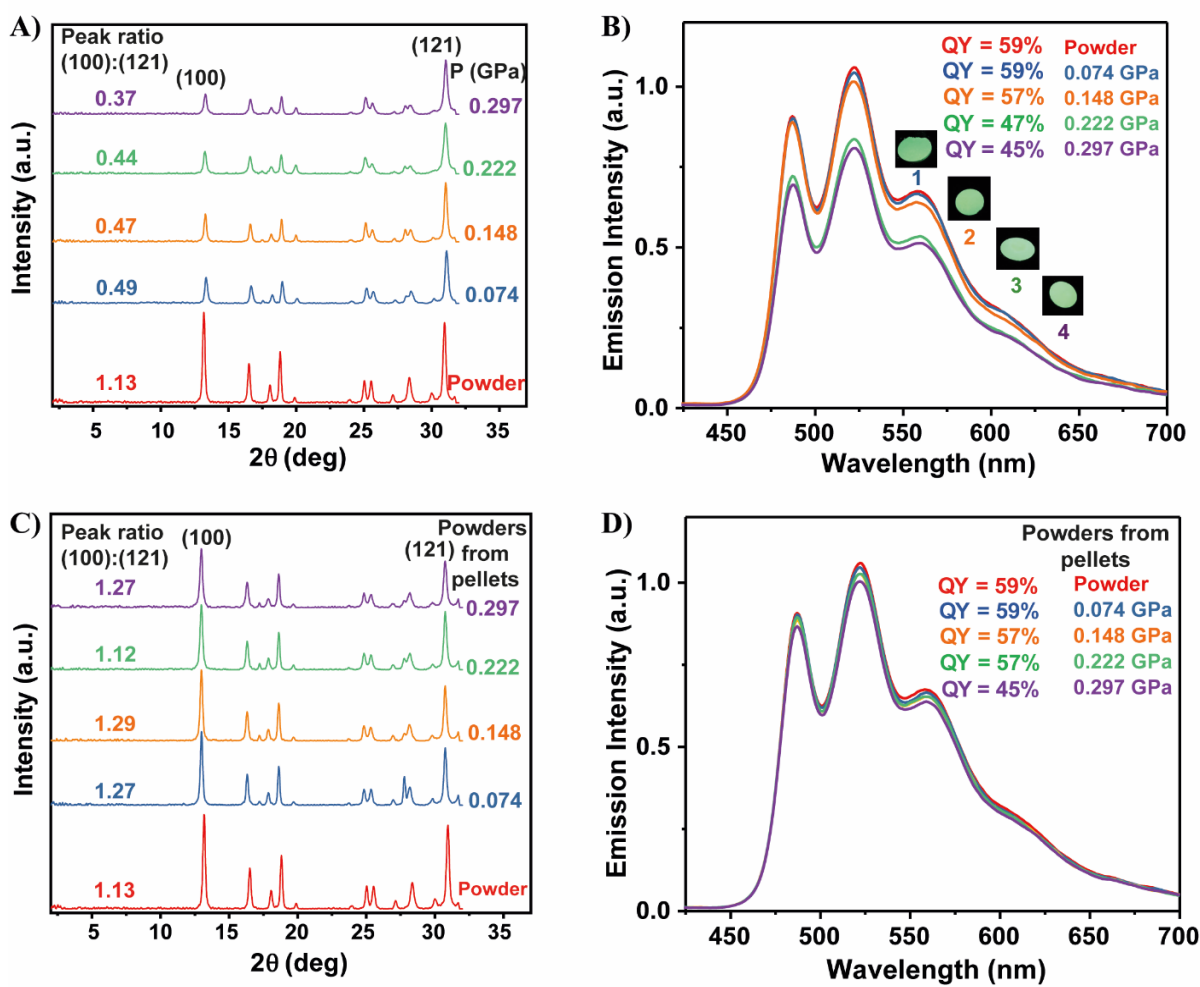

**Figure 4. A-B)** PXRD patterns and emission spectra of the OX-2 pellets compressed at different pressures (indicated as photo inset). **C-D)** PXRD patterns and emission spectra of OX-2 powders recovered after grinding of the former pellets (for simplicity we quoted the pressure values of the pellets).

We discovered that the structural and photophysical changes observed above can be reversed by simply breaking the compressed pellets into a fine powder again (by grinding using a mortar and pestle). The PXRD of the OX-2 powders from the pellets in **Figure 4C** illustrates the recovery of the peak intensity at $2\theta$ ~13° (i.e. (100) plane), comparable to that observed for the as-synthetized OX-2 powder. The data revealed that, although there is a structural deformation of OX-2 under mechanical compression, elastic springback occurs upon grinding to return to its undistorted crystalline structure. What is even more astonishing is the recovery of the emission intensity, especially for the powders recovered from pellets compressed at 0.222 GPa and 0.297 GPa, whose quantum yield changes from 47% and 45% in the pellet form to 57% and 55% in powders, respectively (**Figure 4D**). This exceptional behaviour along with the compressive stress offer further evidence that the LMMCT and LMCT transitions are responsible for the photoluminescence response. It is proposed that when the 2D wave-like layers (Scheme S1B) become distorted under mechanical stress, the distances between the clusters and the ligands are altered, consequently the charge transfer (CT) processes become less efficient and the emission intensity drops. While similar observations have been reported for a number of LMOF materials [14,49], quantum yield recovery as a function of pelleting pressure has not been elucidated.

Extraordinarily, when the compression tests and subsequent grinding step were repeated for a second time on the same recovered material, comparable results were obtained (**Figure S11**). Our results highlight the sensitivity of OX-2 MOF towards an exogenous mechanical stress and the repeatability of this procedure stems from the elastic recovery (springback) of its framework structure from stress relaxation [50], thus offering an exciting avenue for the integration of OX-2 into deformation-based sensors useful for the sensing of mechanical strain stimuli.

*3.4.2 Thermochromic response of OX-2: Temperature-dependent luminescent tests*

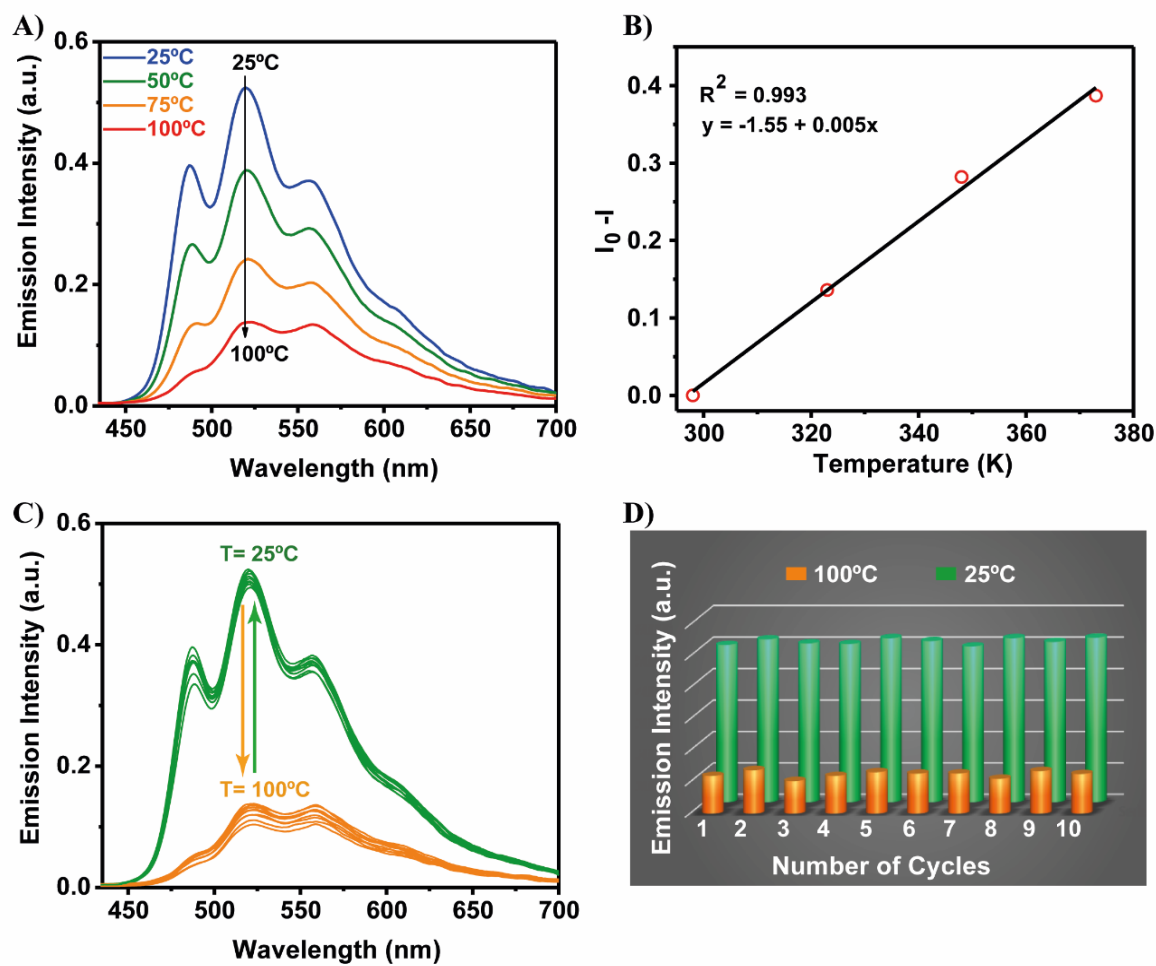

**Figure 5. A)** Emission spectra of OX-2 as a function of temperature. **B)** Correlation between the change in intensity ($I_0$-$I$) vs sample temperature, showing the exceptionally linear response of OX-2 to temperature rise. **C)** Emission spectra of OX-2 recorded at 25 °C and 100 °C subjected to 10 repeated heating-cooling cycles. **D)** Bar chart showing the excellent repeatability and reproducibility of OX-2 for temperature sensing.

The thermochromic ability of OX-2 in detecting changes in temperature has also been studied. Upon increasing the temperature, the emission spectra of OX-2 recorded at 25, 50, 75 and 100 °C revealed a gradual quenching of its emission intensity as shown in **Figure 5A**, resulting in a linear dependence of the intensity change as a function of temperature, i.e. ($I_0$-$I$) vs $T$ ($R^2$ = 0.993, **Figure 5B**). A precisely linear response of a luminescent material to variations in temperature is central to the fabrication of non-invasive luminescent thermometers. However, other parameters such as the reproducibility or repeatability are equally important. Therefore, to test these effects, we have cyclically heated and cooled down the OX-2 powder while studying its emission behaviour.

As shown in **Figure 5C**, when the sample was heated to 100 °C the emission was quenched and when the sample was cooled down to 25 °C the emission was fully recovered. What is most remarkable is that the emission of OX-2 does not exhibit significant losses with each cycle (repeated 10 times, see **Figure 5D**), meaning that the response of this OX-2 MOF is very reproducible under thermal cycling conditions. The above findings indicate that the reason behind this temperature effect should be connected to the structural modification of OX-2 MOF subject to a thermal stimulus. We reason that when the temperature rises, the vibrational modes of the MOFs are also enlarged, causing an increment of non-radiative recombination [13]. Moreover, an alteration of the distances and the coupling between the ligands and the metal clusters will be affected by temperature, as it has been previously observed for other LMOFs [51].

*3.4.3 Luminescent thermometer: Stability tests of pellets and films*

The outstanding abilities of OX-2 to reversibly detect changes in temperature jointly with its easy, scalable, and eco-friendly synthesis make OX-2 a promising candidate for use as a luminescent thermometer. To this end, we have fabricated and examined pellets and films of OX-2 as the ability to shape the material could facilitate the practical implementation of luminescent thermometers. On the one hand, and as explained in the above section, we have fabricated a pellet sample by applying mechanical compression to 200 mg of OX-2 powder under 0.074 GPa, because at this pressure the quantum yield is similar to a pristine powder. On the other hand, homogenous films of OX-2 were prepared by dispersing 20 mg of powder in 20 mL of water under sonication, and then drop casted onto a glass substrate. Both the film and the pellet samples were exposed to a relatively high temperature of 200 °C and 220 °C respectively, using a hot plate. **Figure 6** shows the total disappearance of the emission at higher temperatures, followed by the subsequent recovery of emission upon cooling. Moreover, the response of OX-2 powder, pellets or films to changes in the temperature was demonstrated in the supplementary videos (SI, videos #2-4). Note that because the powder and pellet of OX-2 were placed and maintained on the surface of the hot plate, the emission recovery is slow as this depends on the natural cooling rate of the hot plate (took ~30 mins). However, in the case of the film sample, it was being cyclically placed onto and removed from the hot plate surface, and thus, a visual inspection of supplementary video #4 shows a faster response demonstrating the reversibility of OX-2 to a transient temperature stimulus.

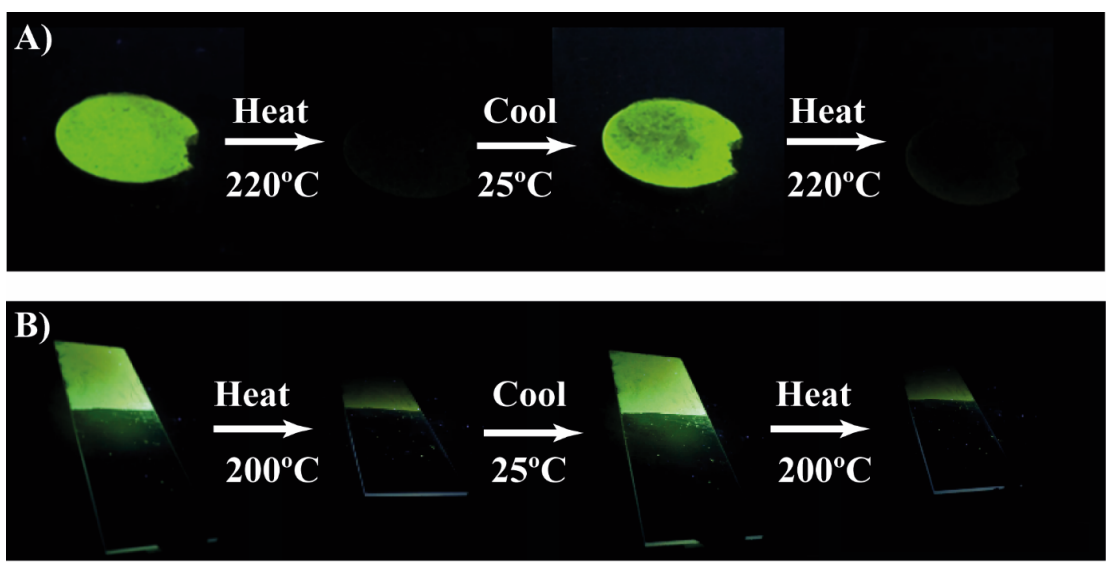

**Figure 6.** Photographs of the OX-2 pellet (A) and film (B) upon cyclic exposures to the high (200-220 °C) and room (25 °C) temperature conditions. The heat source was a digitally controlled hot plate positioned at the bottom surface of the samples (not visible in photographs taken in the dark).

**3.5 Electroluminescence of OX-2 for solid-state LEDs**

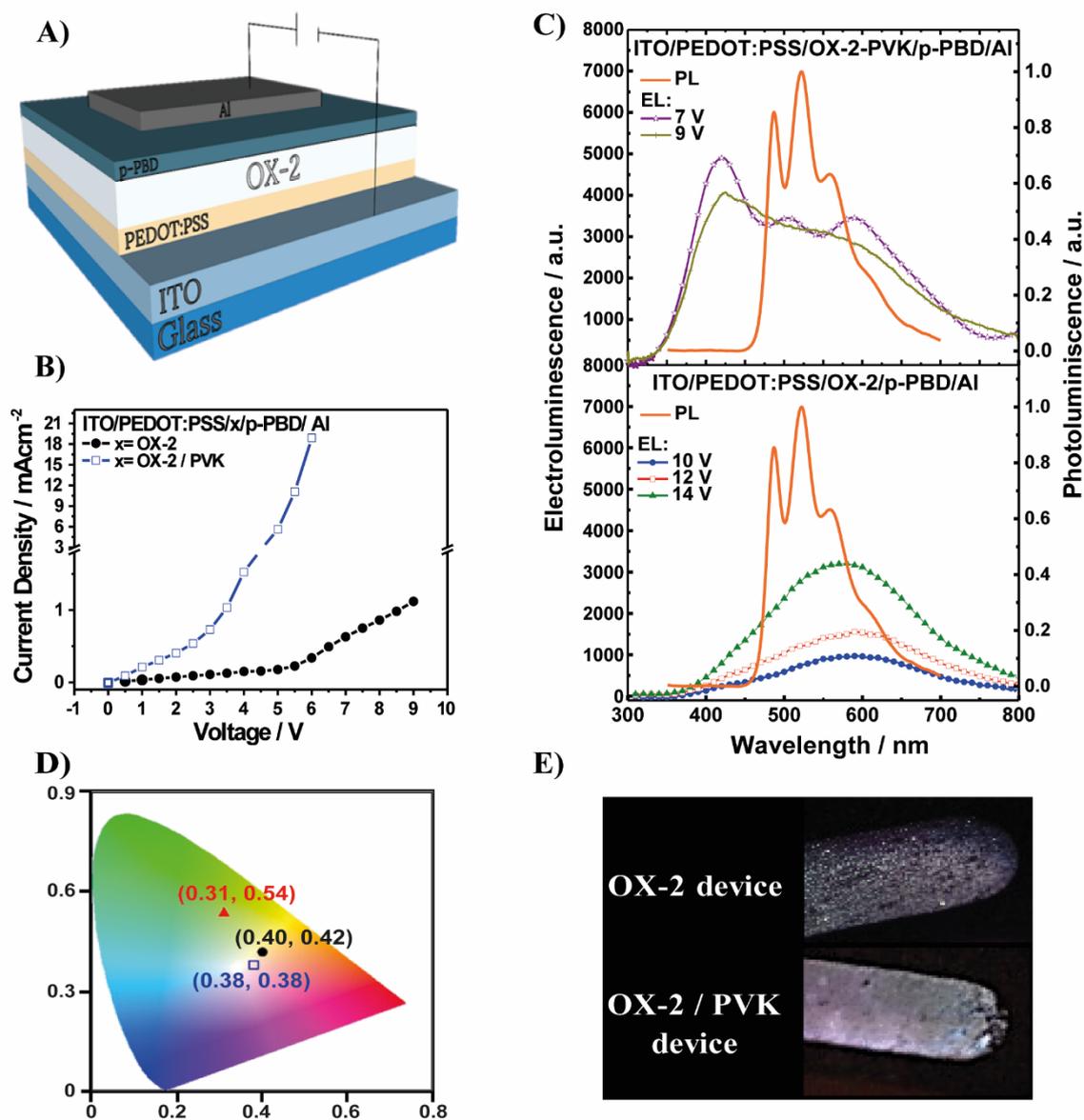

**Figure 7. A)** Schematic representation of the LED device architecture, where the OX-2 MOF acts as the electroluminescent (EL) layer. **B)** Comparison of the current versus voltage curves and **C)** their corresponding electroluminescent spectra of the MOF-LED devices (OX-2 - black, and OX-2 / PVK - blue). The grey line is the photoluminescence (PL) spectrum of the OX-2 powder. **D)** Graph displaying the CIE coordinates corresponding to the photoluminescence spectrum of OX-2 (red triangle) and the electroluminescence of the different MOF-LEDs (OX-2 – black dot, and OX-2 / PVK – blue square). **E)** Photo of the turn-on EL devices incorporating OX-2 (upper photo) and OX-2 / PVK (bottom), showing the whitish emission.

To further explore all the possible applications of this material, OX-2$_w$ was implemented as an electroluminescent layer in a bottom-up LED device using a transparent anode electrode,

ITO (work function, 4.7 eV) and an aluminium cathode (work function, 4.3 eV) [52]. However, this single-layer device configuration did not show any stable electroluminescence and this could be related to 2 main factors: (i) the energy levels of the electrodes and the electroluminescent material do not match, and/or (ii) the carrier mobility is not balanced in the device [53]. Hence, additional layers such as PEDOT:PSS and 2-(4-tert-butylphenyl)-5-(4-biphenylyl)-1,3,4-oxadiazole (p-PBD) were implemented in the device to overcome the above problems (**Figure 7A**). This approach gives a stable working device with a turn-on voltage of around 5 V (**Figure 7B**). A visual inspection of this device under 11 V shows an inhomogeneous white-light emission (**Figure 7E**) with a CIE coordinates of (0.40, 0.42) (**Figure 7D**). This whitish emission is the result of a very broad electroluminescence spectra (EL) which ranges from 400 nm to 800 nm (**Figure 7C**), suggesting that the inhomogeneity of the EL LED is connected to different emissive species. The features of the spectra and the maximum intensity wavelength (~590 nm, 6990 cm$^{-1}$) is independent of the forward driving voltage, however, the EL spectra is not correlated to the vibrationally resolved photoluminescent spectra (**Figure 3D** and **Figure 7C**) and its CIE coordinates (0.31, 0.54).

To determine the electroluminescence behaviour and the different origin of photo- and electro-luminescence species, a deconvolution of the obtained spectra was performed using a Gaussian fit. Two different bands were required to satisfy the fit, one centred at 520 nm and the other at 620 nm (Figure S12A). The band at 520 nm corresponds to the maximum of the photoluminescence spectra of the OX-2, even when the vibrational structure is not observed in the EL spectra; it is possible to attribute this emissive species to the same one observed in the photoluminescence. The lack of vibrational resolution in the spectrum feature could be explained by the heterogeneous emissive layer and the weak electroluminescent signal. Furthermore, the origin of the 620 nm species is only caused by the electric field when the electron-hole recombination occurs in the device. Previous studies have shown that the presence of the red emitter species in the EL is mainly due to electron-hole recombination in a trap state found in the emissive layer. Among all the possibilities, the most frequently observed factors are attributed to grain boundaries between the nanocrystals [54], or the unbalanced charge carrier mobility (which is usually induced by the presence of deep electronic trap states in the studied materials) [35].

With the aim of reducing the grain boundaries and at the same time optimizing the charge carrier mobility, the OX-2 was incorporated in a semiconducting host matrix forming a hybrid organic-inorganic LED. The selection of the polyvinyl carbazole (PVK), a hole transport polymer, was guided by the conclusions of several recent examples in the literature [32,33,35]. There, it was highlighted that the majority charge carriers through silver cluster to silver clusters are electrons leading to an unbalanced electron-hole carrier [36,55]. Those experimental observations were supported by the calculated electronic energy levels for the HOMO and LUMO of the silver clusters [42,56,57], where their LUMO is accessible independent of the cathode used, meanwhile the HOMO level is difficult to be reached. **Figure 7B** shows that the turn-on voltage of this device is now 1.5 V (~3 times lower in comparison with the device without PVK) and displays a homogenous white light emission (**Figure 7E**) with CIE coordinates (0.38, 0.38) (**Figure 7D**). In this case, the EL spectrum is much more intense (by five folds) at a lower voltage of 7 V (versus 10 V), indicating that the developed approach to enhance the hole mobility is effective for improving the LED performance. Surprisingly, in this case the EL spectrum is changing at different driving voltages leading to more complicated electroluminescent behaviour.

To shed light on the above experimental observations, Gaussian deconvolution was performed and three different emission bands were obtained (**Figure S12B**). Although two of them, the bands at 520 nm and 620 nm, have been obtained and described for the previous OX-2 device, the new one centred at 420 nm is strongly contributing to the EL spectra. This specie is far from the first vibrational band of the photoluminescence spectrum (485 nm) but it does not differ from the EL observed in a PVK pristine polymer device using the same conditions [35]. Therefore, this specie might correspond to the S1-excimer emission from PVK, which is further supported by the decrease in its intensity observed at higher voltage (9 V), where the electromer specie (band at ~610 nm) is formed at its expense (**Figure S12B**) [58]. The 520 nm specie has the same origin of the PL emission as it has been explained above, however in this case the intensity has been increased in comparison to the OX-2 device without PVK, indicating a better device performance. The emission at 620 nm is also enhanced, but in this instance, the higher emission is more likely originating from the electromer emission (~610 nm) than a favoured formation of the red-emitting specie.

Based on these results, the electroluminescent behaviour might be explained by the following mechanism: the electrons can be injected to the LUMO of the silver clusters and/or to the PVK LUMO, meanwhile the holes are injected preferentially to the PVK HOMO, as its energetic level is more accessible. The silver clusters of the OX-2 act as a conductive pathway for both charges which can hop between them. In some cases, the electron/hole is recombined from the conduction-valence band of the PVK giving an emission at 420 nm, and in other cases, they are recombined in the OX-2 with an emission at 520 nm. The emission at 620 nm, detected even when the polymer is absent, suggests that the holes are more likely to be trapped in the defects of the MOF framework (previously observed for similar materials) [32,35,36] than in the grain boundaries as not a lot of differences were observed in the emission band with more homogenous polymer film. In more detail, the electrons injected (through the polymer or the cluster) will hop from one silver cluster to the next silver cluster until they meet a hole where emissive recombination can occur. Such an electron hopping mechanism between the silver clusters was previously reported for other materials with silver clusters, where the Ag bond distance was less than 2 Å [59]. Importantly, the EL behaviour observed above further supports the role played by the LMCT and/or LMMCT processes to enable the emission of excited OX-2 proposed above (section 3.2).

## 4. Conclusions

In summary, we have developed four synthetic approaches for the fabrication of a silver-based OX-2 MOF. We established that the OX-2 synthesized in water has the highest quantum yield of ~60% in the solid state, which we ascribed to the hybridization between the Ag atoms of the MOF and the oxygen orbitals of guest water molecules. This is an enormous advantage as the cost-effective synthesis of OX-2 has been successfully scaled up employing an eco-friendly solvent (water) and mild conditions (room temperature). Moreover, OX-2 MOF has proven to be a very robust material.

We discovered that OX-2 MOF is mechanochromic and thermochromic, exhibiting a unique luminescent response subject to mechanical stress and changes in temperature. Its thermochromic behaviour has a linear response of the emission intensity with temperature, which was established to be very reproducible and repeatable to enable luminescent thermometry. Moreover, we show that OX-2 MOF is a good electroluminescent material as their excited state mechanism is based on LMCT and LMMCT processes. Subsequently, we have designed and fabricated a MOF-LED prototype device integrating OX-2 as the electroluminescent layer. This study exemplifies OX-2 as one of the first examples in the emerging field of MOF-LEDs. Our work marks the first foray into the silver-based luminescent frameworks, targeting technological applications in the field of photonic sensors and LED lightings.

**Supplementary Information**

Supplementary Information to this article can be found online at http:// .

**Competing financial interests**

M.G. and J.C.T. have financial interests in the materials technology described in this study, through patenting of the Ag-MOFs for commercial exploitation.

**Funding**

This work was supported by the EPSRC IAA award (EP/R511742/1) and the ERC Consolidator Grant PROMOFS (through the grant agreement 771575). This work was also supported by the Research Foundation-Flanders (FWO, grant nos. G.0B39.15 and G098319N); FWO postdoctoral fellowships to C.M. (grant nos. 12J1719N and 12J1716N), the KU Leuven Research Fund (C14/15/053); the Flemish government through long-term structural funding Methusalem (CASAS2, Meth/15/04); the Hercules Foundation (HER/11/14); and the Belgian Federal Science Policy Office (IAP-VII/05).

**Acknowledgements**

We thank the Research Complex at Harwell (RCaH) for access to the Nicolet iS10 FTIR spectrometer. We thank Dr Jennifer Holter for assisting in the FESEM-EDX measurements.

# *Supplementary Information*

# *for*

# Highly luminescent silver-based MOFs: Scalable eco-friendly synthesis paving the way for photonics sensors and electroluminescent devices

Mario Gutiérrez,[1] Cristina Martín,[2,3] Barbara E. Souza,[1] Mark Van der Auweraer,[3] Johan Hofkens,[3] and Jin-Chong Tan[1*]

[1]*Multifunctional Materials & Composites (MMC) Laboratory Department of Engineering Science, University of Oxford, Parks Road, Oxford OX1 3PJ, United Kingdom.*

[2]*Unidad de Medicina Molecular, Centro Regional de Investigaciones Biomédicas, Albacete, Spain.*

[3]*Molecular Imaging and Photonics, Department of Chemistry, Katholieke Universiteit Leuven, Celestijnenlaan 200F, 3001 Leuven, Belgium*

*Correspondence to:*  jin-chong.tan@eng.ox.ac.uk

**Scheme S1.** Schematic representation of the **(A)** binuclear Ag$_2$(BDC)$_2$ building units (CCDC 198096) [1] and (B) the interweaved 2D layers giving rise to the dense 3D framework of OX-2 MOF.

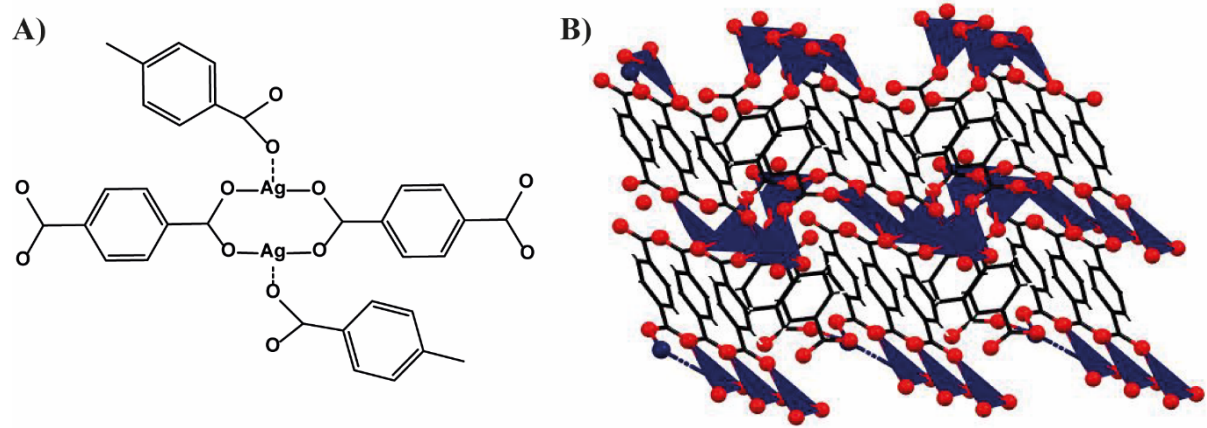

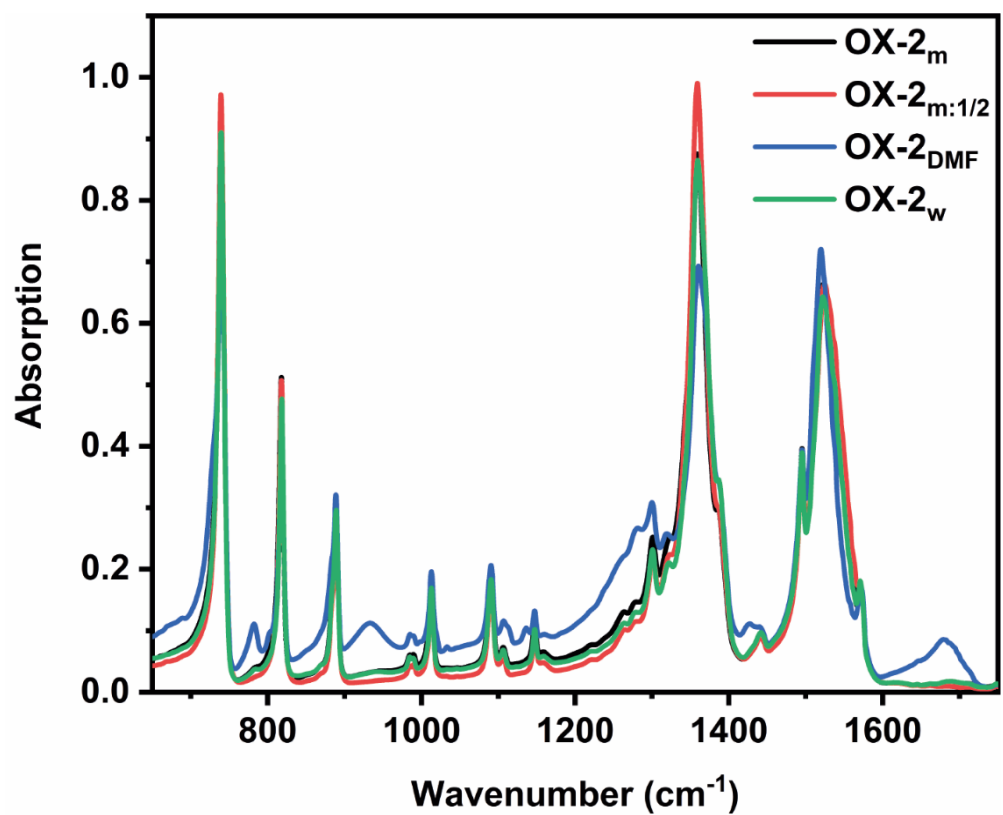

**Figure S1.** FTIR spectra of OX-2$_m$ (black line), OX-2$_{m:1/2}$ (red line), OX-2$_{DMF}$ (blue line) and OX-2$_w$ (green line).

**Figure S2.** FESEM micrographs of (A) OX-2$_m$ and (B) OX-2$_{DMF}$ MOFs.

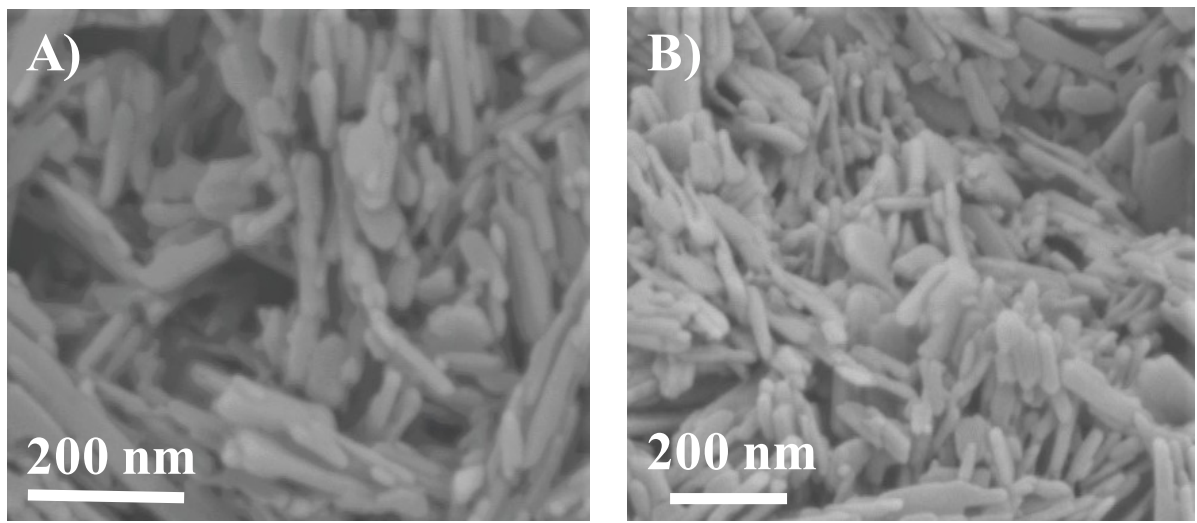

**Figure S3.** (A-B) AFM topography images of the elongated nanoplates of OX-2$_w$. (C-D) Thickness profiles of the OX-2$_w$ elongated nanoplates obtained from the AFM image in (A) and (B) respectively, showing a thickness of tens of nm.

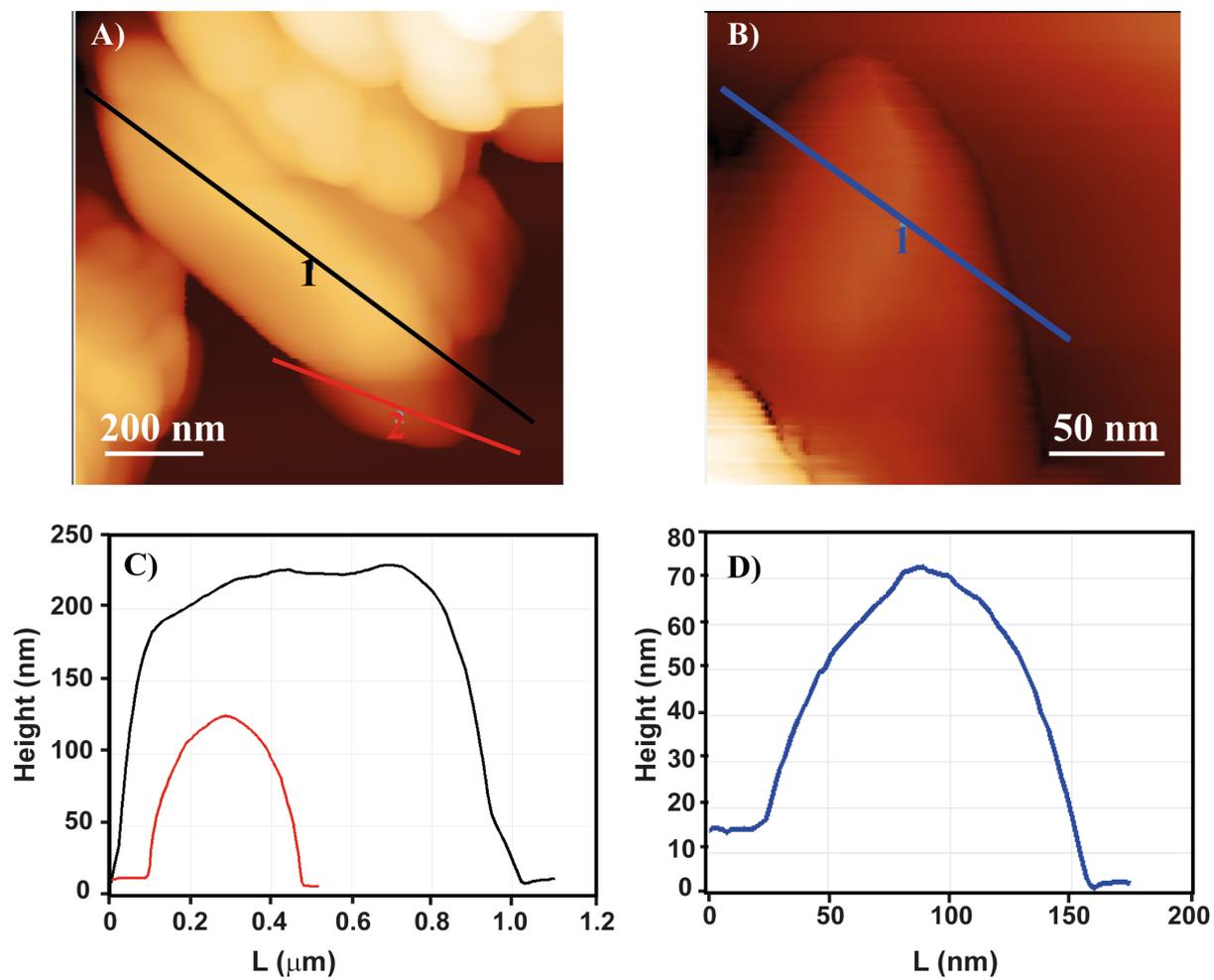

**Figure S4.** FESEM-EDX micrographs of (A-C) OX-2$_{m:1/2}$ and (D-F) OX-2$_w$, showing the "flower" shaped crystals and the homogeneously distributed elongated nanoplates, respectively.

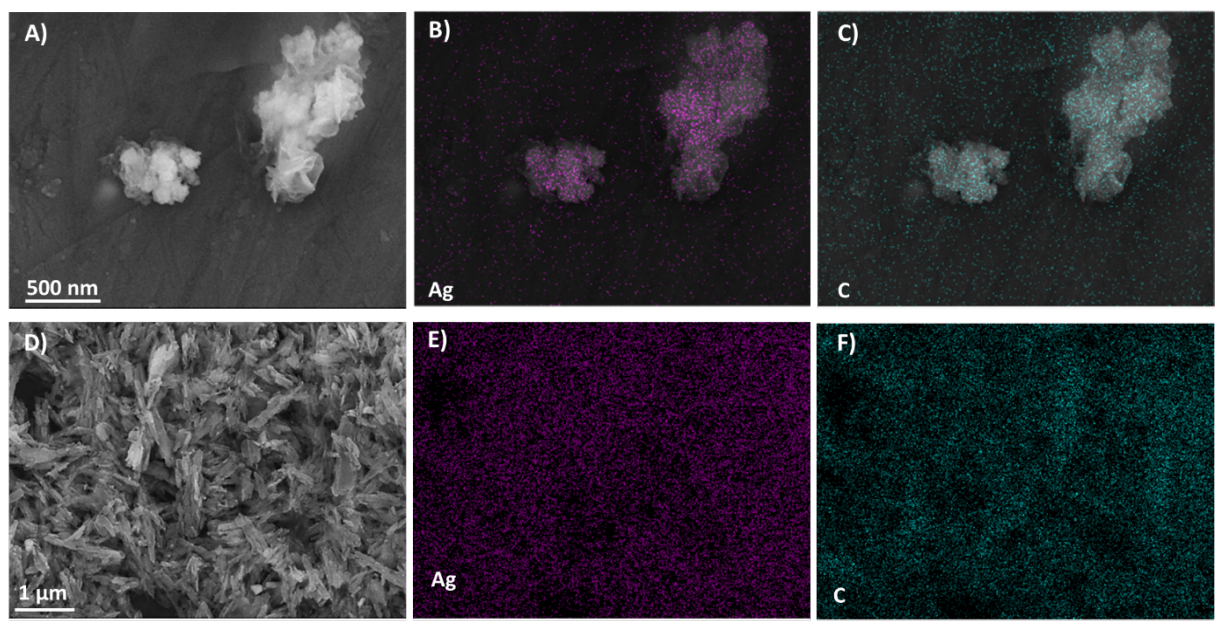

**Table S1.** Representation of the %weight and %atomic of C, Ag and O elements estimated from the EDX maps corresponding to the FESEM micrographs.

| Materials (micrograph) | Element | % weight | % atomic |
|---|---|---|---|
| OX-2$_{m:1/2}$<br>**Figures 2 E-G** | C<br>Ag<br>O | 23<br>5<br>5 | 41<br>1<br>6 |
| OX-2$_{m:1/2}$<br>Micron-sized "columns"<br>**Figures 2 H-J** | C<br>Ag<br>O | 33<br>0.43<br>3.5 | 52<br>0.08<br>4 |
| OX-2$_{m:1/2}$<br>"Flower"-shaped crystals<br>**Figures S4 A-C** | C<br>Ag<br>O | 16<br>5<br>4 | 30<br>1<br>6 |
| OX-2$_w$<br>Elongated nanoplates<br>**Figures S4 D-F** | C<br>Ag<br>O | 40<br>22<br>6 | 66<br>4<br>8 |

**Figure S5.** Nanosecond-picosecond emission decays of (A) BDC linker, (B) OX-2$_{DMF}$, (C) OX-2$_m$, (D) OX-2$_{m:1/2}$ and (E) OX-2$_w$ in solid-state. The observation wavelengths are indicated as inset and the samples were excited at 365 nm. The solid lines are from the best-fit using a multiexponential function.

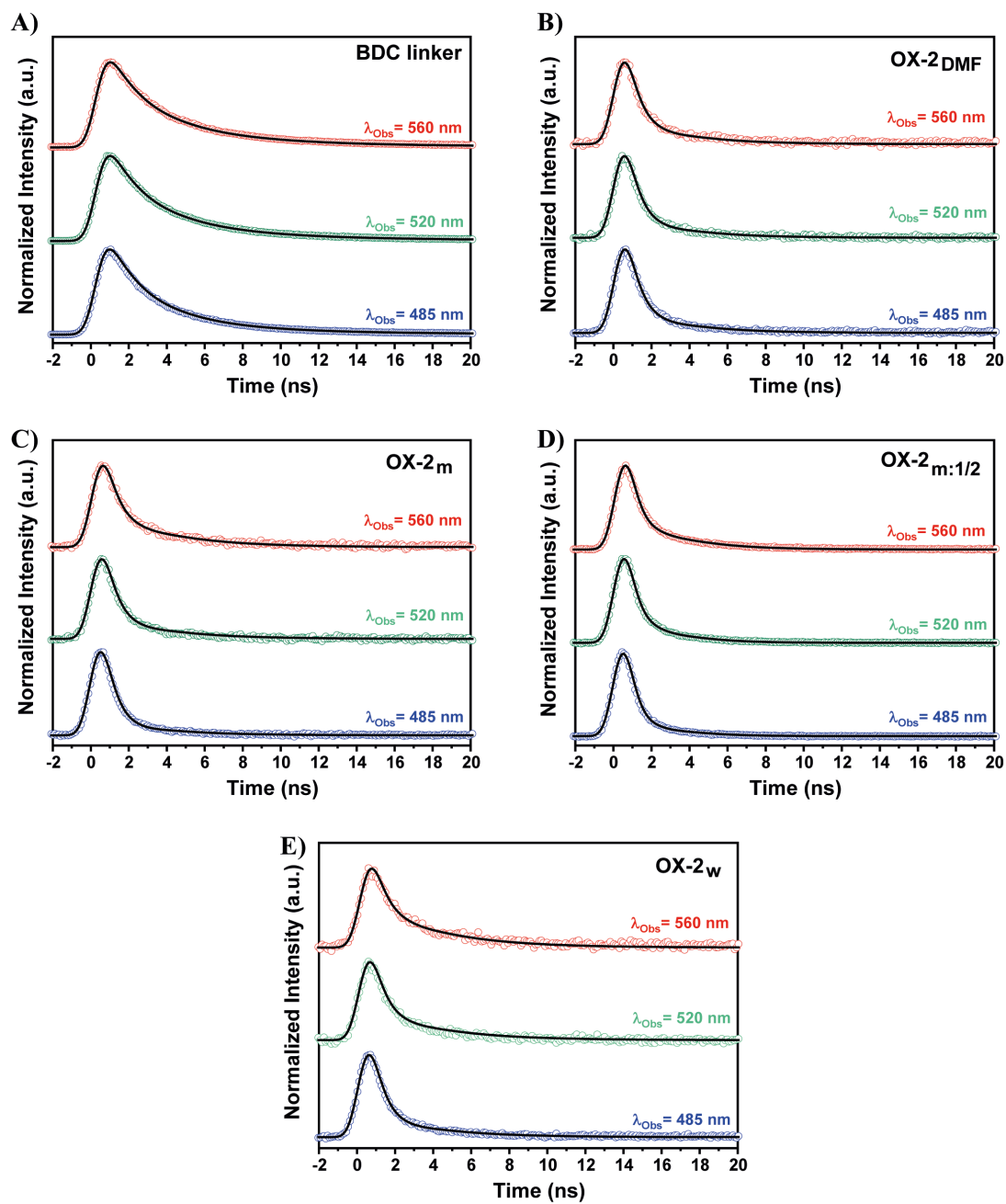

**Table S2.** Values of time constants ($\tau_i$), pre-exponential factors ($a_i$) and normalized (to 100) fractional contributions ($c_i = \tau_i a_i$) obtained from the fit of the emission decays of the BDC linker in solid-state upon excitation at 365 nm and for the observation wavelength ($\lambda$ obs) indicated.

| Sample | $\lambda$ obs / nm | $\tau_1$ / ns | $a_1$ | $c_1$ | $\tau_2$ / ns | $a_2$ | $c_2$ | $\tau_3$ / ns | $a_3$ | $c_3$ |
|---|---|---|---|---|---|---|---|---|---|---|
| | 485 | 1.7 | 0.080 | 41 | 4.2 | 0.037 | 46 | 16.5 | 0.002 | 12 |
| BDC | 520 | 1.7 | 0.074 | 35 | 4.2 | 0.042 | 49 | 16.5 | 0.004 | 16 |
| | 560 | 1.7 | 0.069 | 32 | 4.2 | 0.044 | 50 | 16.5 | 0.004 | 18 |

**Table S3.** Values of time constants ($\tau_i$), pre-exponential factors ($a_i$), and normalized (to 100) fractional contributions ($c_i = \tau_i a_i$) obtained from the fit of the emission decays of OX-2$_m$, OX-2$_{m:1/2}$, OX-2$_{DMF}$ and OX-2$_w$ in solid-state upon excitation at 365 nm and observation as indicated.

| Sample | $\lambda$ obs / nm | $\tau_1$ / ps | $a_1$ | $c_1$ | $\tau_2$ / ns | $a_2$ | $c_2$ |
|---|---|---|---|---|---|---|---|
| | 485 | 540 | 0.151 | 73 | 3.4 | 0.09 | 27 |
| OX-2$_m$ | 520 | 540 | 0.098 | 57 | 3.4 | 0.012 | 43 |
| | 560 | 540 | 0.080 | 42 | 3.4 | 0.017 | 58 |
| | 485 | 350 | 0.103 | 67 | 2.0 | 0.009 | 33 |
| OX-2$_{m:1/2}$ | 520 | 350 | 0.094 | 54 | 2.0 | 0.014 | 46 |
| | 560 | 350 | 0.090 | 52 | 2.0 | 0.014 | 48 |
| | 485 | 540 | 0.049 | 61 | 3.0 | 0.006 | 39 |
| OX-2$_{DMF}$ | 520 | 540 | 0.051 | 61 | 3.0 | 0.006 | 39 |
| | 560 | 540 | 0.048 | 57 | 3.0 | 0.006 | 43 |
| | 485 | 540 | 0.126 | 58 | 3.5 | 0.014 | 42 |
| OX-2$_w$ | 520 | 540 | 0.066 | 44 | 3.5 | 0.013 | 56 |
| | 560 | 540 | 0.047 | 32 | 3.5 | 0.016 | 68 |

**Figure S6.** Comparison of the PXRD patterns of the [Ag-(BDC)$_{1/2}$]$_n$ reported [1] and those of OX-2$_w$ MOFs obtained through a small scale synthesis (hundreds of mg) and a scale-up synthesis (10 g).

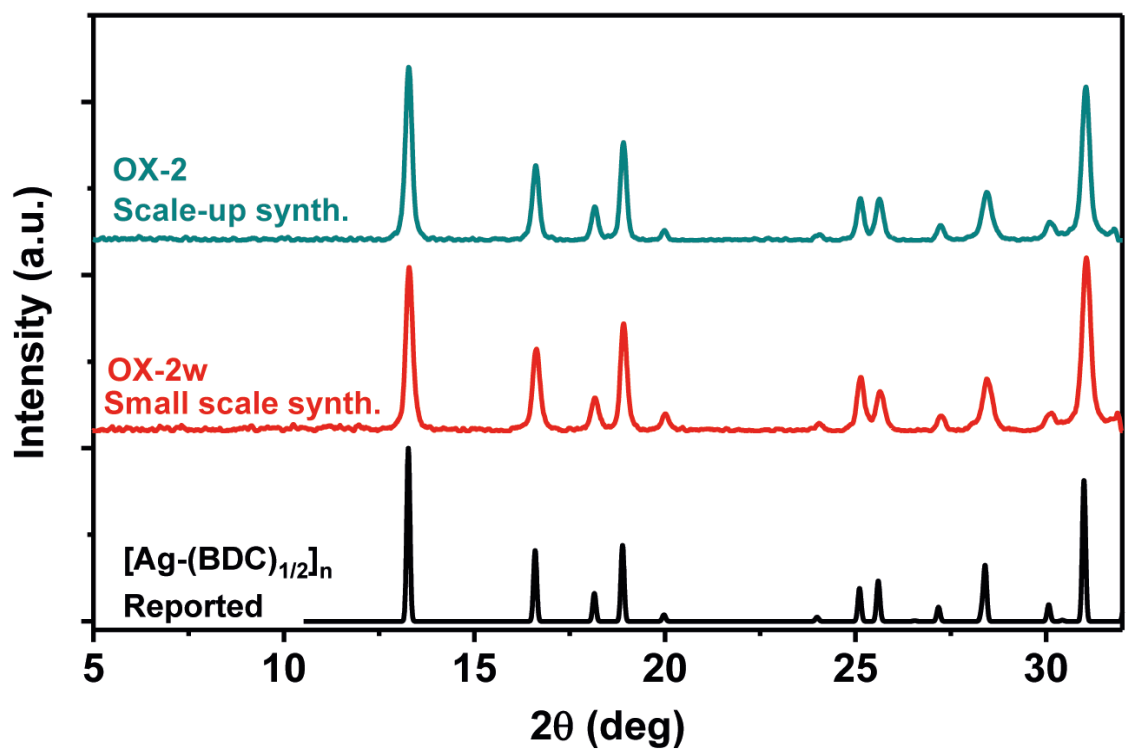

**Figure S7.** Excitation-emission map of the OX-2$_w$ obtained through our scale-up method.

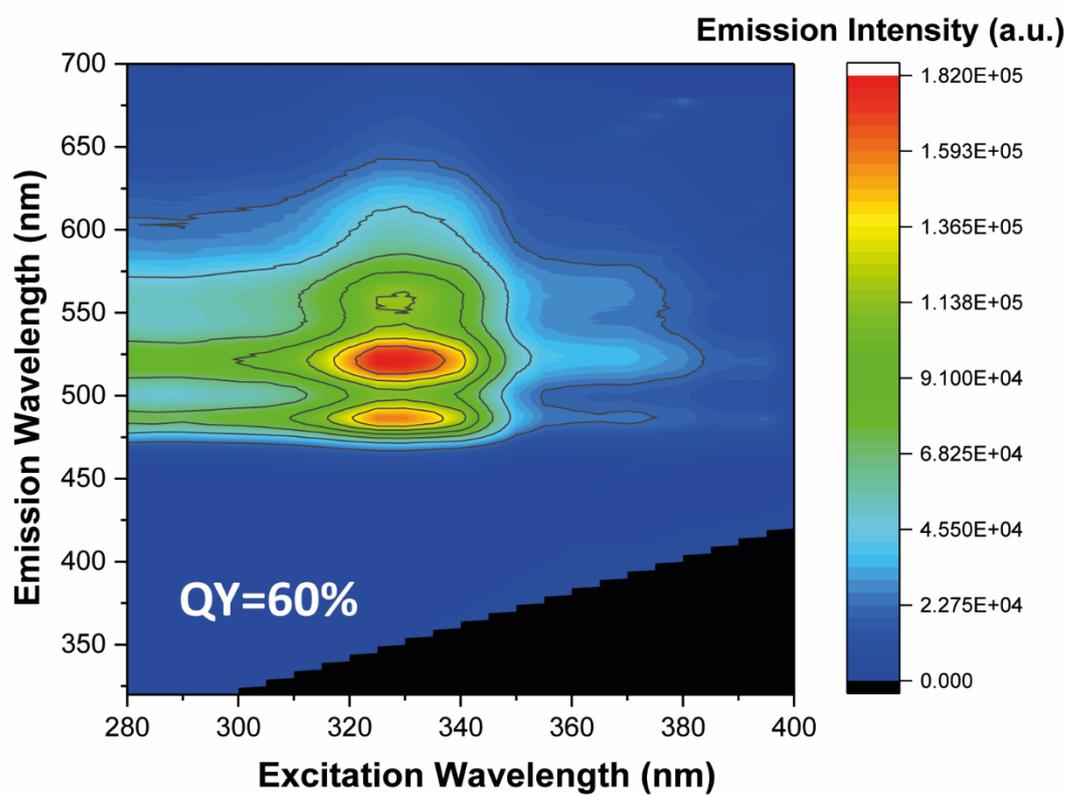

**Figure S8**. PXRD patterns of OX-2$_w$ after being soaked in water for different time intervals, ranging from 1 to 21 days.

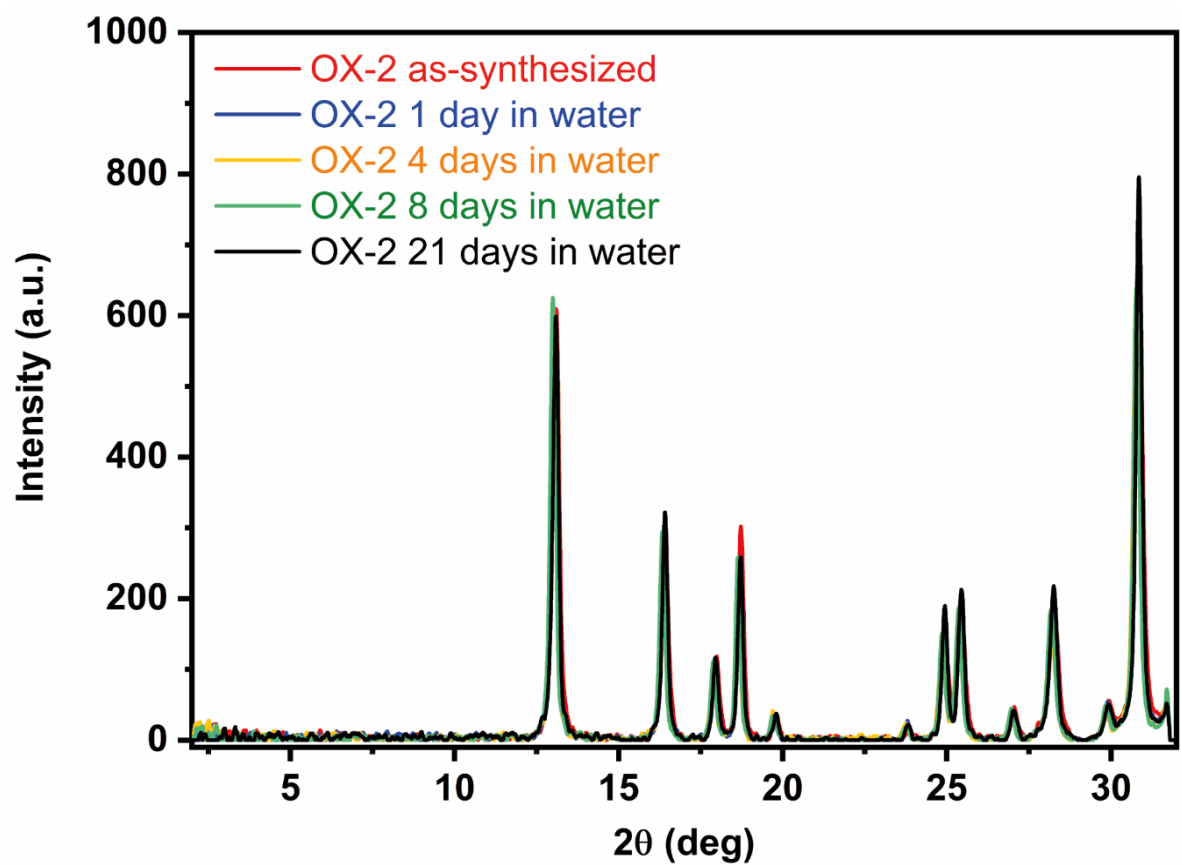

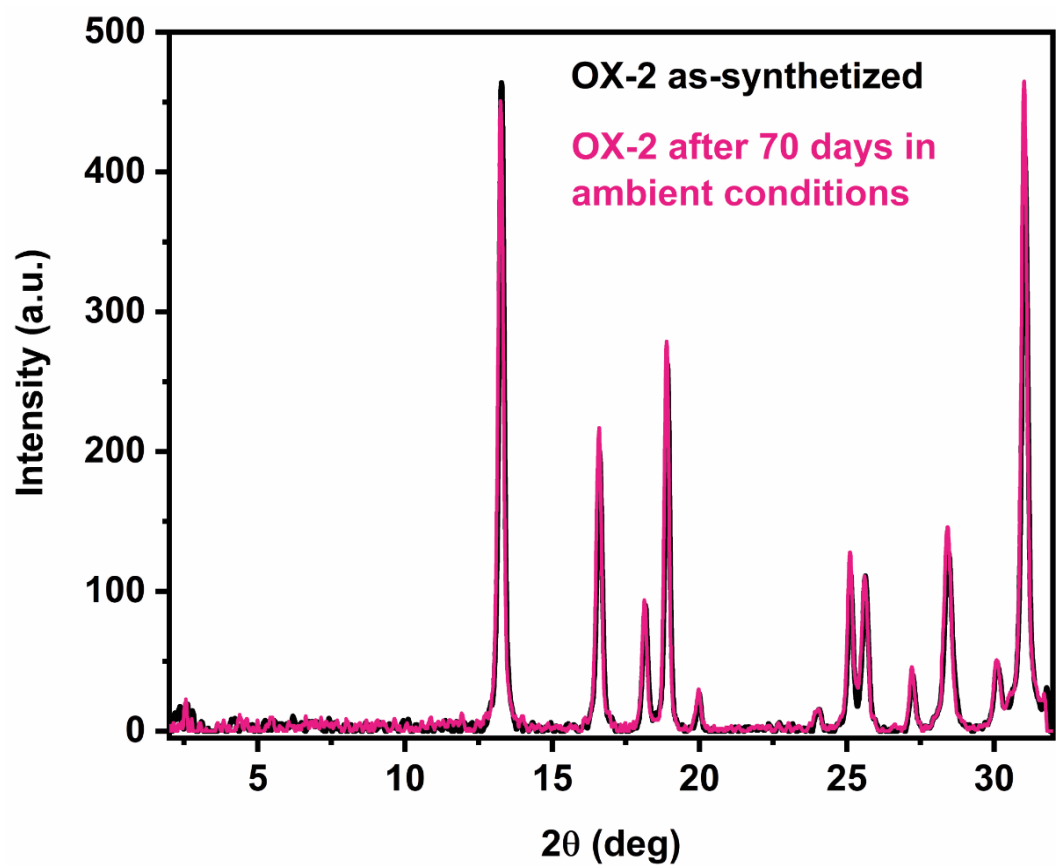

**Figure S9**. PXRD patterns of OX-2$_w$ MOF as-synthetized and after being exposed to ambient conditions in the lab (day light, ~40% humidity, etc) for 70 days.

**Figure S10.** Emission of spectra of OX-2$_w$ MOF before and after being exposed to ambient conditions in the laboratory (day light, ~40% humidity, etc) for up to 70 days. The inset shows a minimal decrease in the quantum yield values from 60% to 57%.

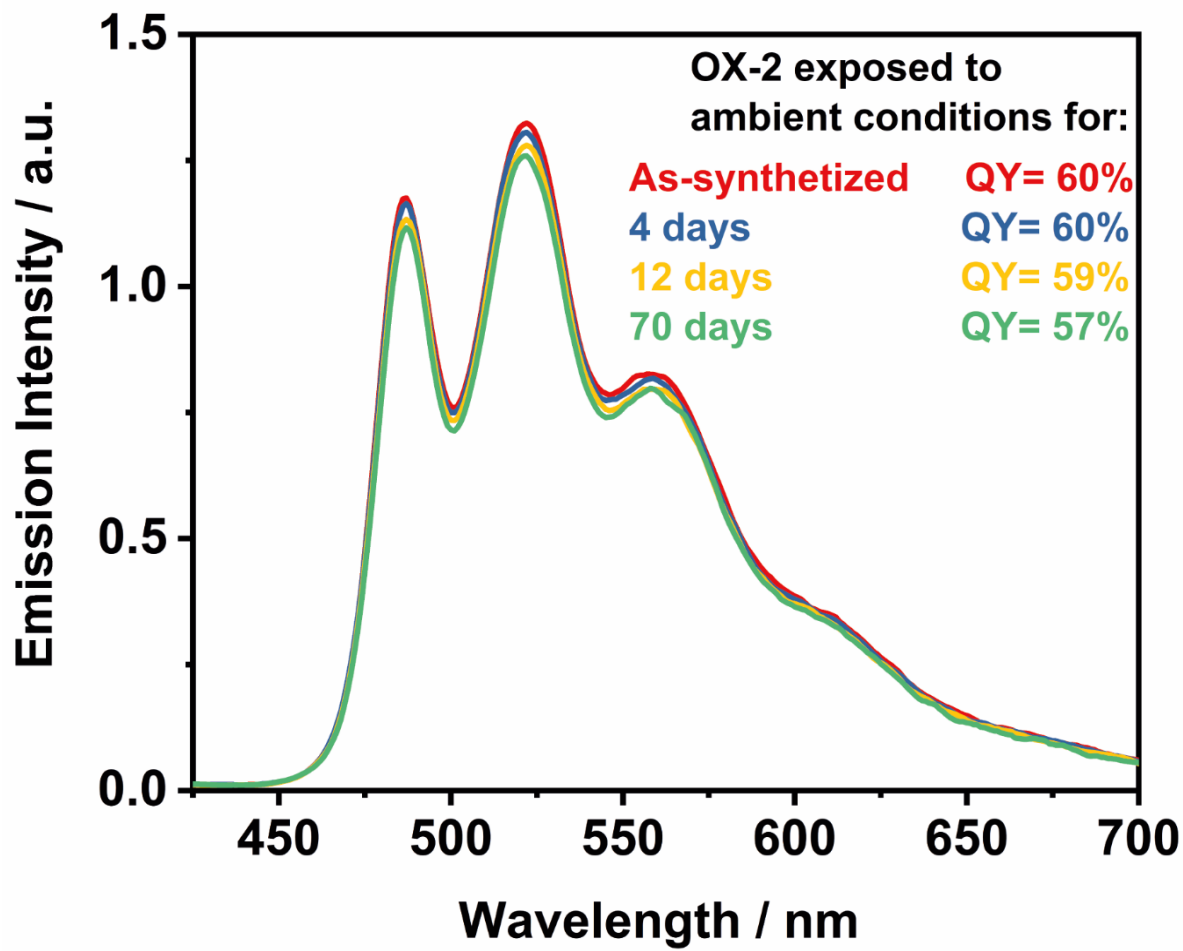

**Figure S11.** (A-B) PXRD patterns and emission spectra of OX-2 pellets compressed at different pressures for the second time. (C-D) PXRD patterns and emission spectra of OX-2 powders obtaining by grinding the former pellets.

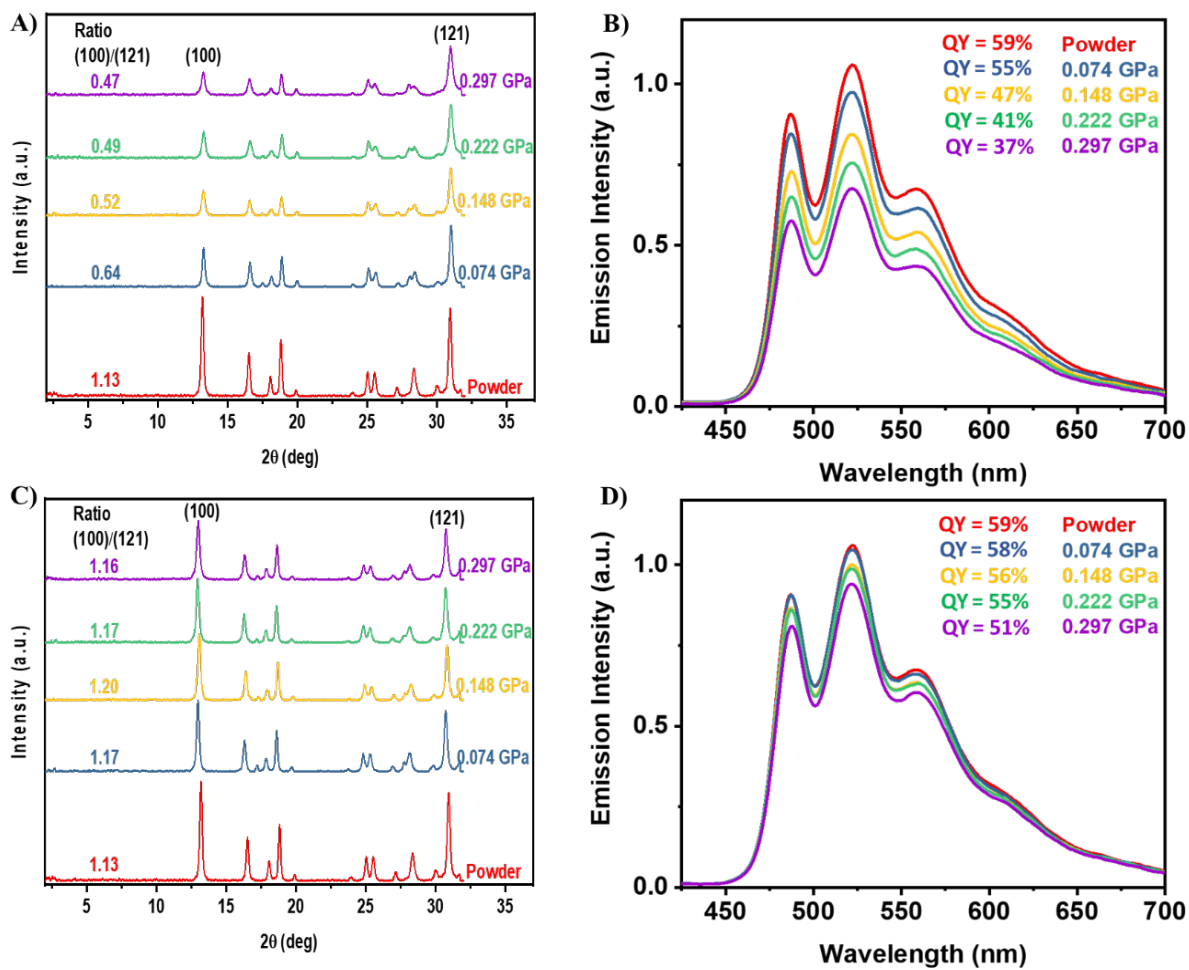

**Figure S12.** Deconvoluted electroluminescence spectra and their fits of the (A) OX-2 and (B) OX-2 / PVK LED devices at different applied voltages (indicated as inset).

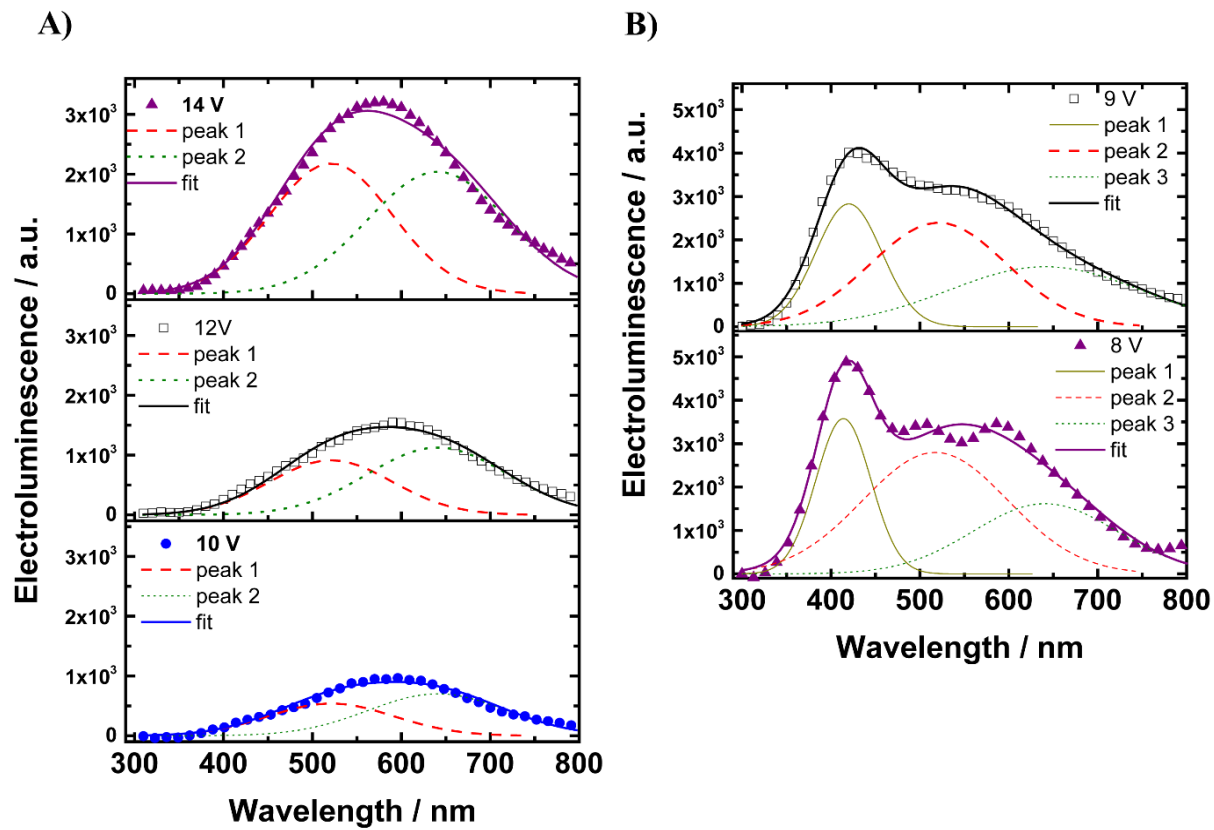

**Leaching Experiments**

To test the chemical stability of OX-2$_w$, we have performed leaching experiments in order to quantify the amount of BDC linkers being released from the framework over time. To this end, we have poured 500 mg of OX-2$_w$ in 30 mL of water for a few days. During those days, the UV-vis absorption spectra of some aliquots of the supernatant were recorded to estimate the amount of BDC linker released into the solution.

As a reference and for comparing the chemical decomposition of OX-2$_w$, we have prepared a solution of 11 mg of BDC linker in 30 mL of water which corresponds to a 5% decomposition of the MOF. This calculation was made by taking into account the following steps:

- The chemical formula of OX-2 MOF is [Ag$_2$BDC]$_n$, which gives a total molecular weight of 380 g/mol. From there, 216 g/mol (56%) corresponds to Ag$_2$ atoms, while 164 g/mol (44%) to the BDC linker.

- This means that in the prepared solution (500 mg of OX-2 in 30mL water), 220 mg of BDC molecules (500 mg x 0.44) would be released if the MOF completely degraded.

- Therefore, for an estimated MOF decomposition of 5%, a solution of 11 mg of BDC linker was dissolved in 30 mL of water.

The UV-vis absorption spectra of the different aliquots of the supernatant collected from the OX-2$_w$ suspension over time (see Figure S13) compared to the absorption of the prepared BDC solution (simulating a 5% decomposition) is ~45 times lower, the data reveal that the degradation of OX-2$_w$ in water is negligible (≤0.1%) over this time.

**Figure S13.** Absorption spectra of the BDC solution prepared for simulating a 5% of MOF degradation and the different aliquots of the supernatant collected over time from a suspension of 500 mg of OX-2 in 30 mL of water.

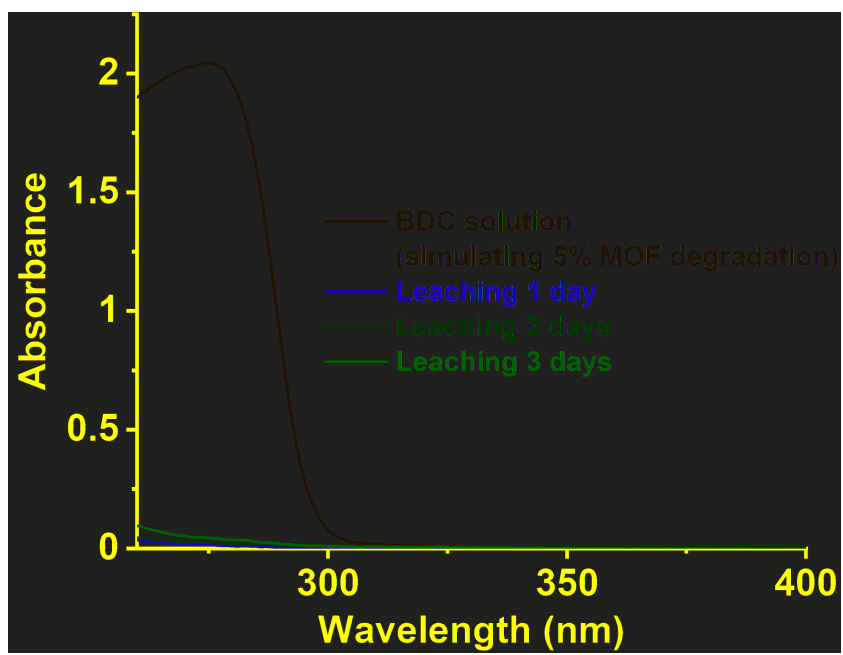